\documentclass{acm_proc_article-sp} 

\usepackage{fainekos-macros}
\usepackage{color}
\usepackage[english]{babel}
\usepackage{acronym}
\usepackage{url}
\usepackage[ruled]{algorithm}
\usepackage{algpseudocode}
\usepackage{paralist}
\usepackage{ragged2e}
\setboolean{HIGHLIGHTCHANGES}{FALSE}

\newcommand{\slabel}{\sigma}

\newcommand{\labelSet}{\Sigma}

\newcommand{\trans}[1]{\xrightarrow{#1}}

\newcommand{\simu}{\mathcal{S}}
\newcommand{\bisimu}{\mathcal{B}}
\newcommand{\Ft}{\Fc_t}
\newcommand{\Fd}{\Fc_d}
\newcommand{\partition}{\mathcal{P}}
\newcommand{\SHS}{\Sigma}

\newcommand{\egm}{s}
\acrodef{NSR}{Normal Sinus Rhythm}
\acrodef{AP}{Action Potential}
\acrodef{ICD}{Implantable Cardioverter Defibrillator}
\acrodef{SVT}{SupraVentricular Tachycardia}
\acrodef{VT}{Ventricular Tachycardia}
\acrodef{VF}{Ventricular Fibrillation}
\acrodef{SVT}{SupraVentricular Tachycardia}
\acrodef{EGM}{electrogram}
\acrodef{CA}{cellular automata}

\newboolean{REPORT}
\setboolean{REPORT}{FALSE}
\begin{document}

\title{Model Checking Implantable Cardioverter Defibrillators}

\author{
\alignauthor
Houssam Abbas, Kuk Jin Jang, Zhihao Jiang, Rahul Mangharam\\
       \affaddr{Department of Electrical and Systems Engineering}\\
       \affaddr{University of Pennsylvania, Philadelphia, PA, USA}\\
       \email{\{habbas,  jangkj, zhihaoj, rahulm\}@seas.upenn.edu}
}

\maketitle
\begin{abstract}
Ventricular Fibrillation is a disorganized electrical excitation of the heart that results in inadequate blood flow to the body.
It usually ends in death within seconds.
The most common way to treat the symptoms of fibrillation is to implant a medical device, known as an \emph{Implantable Cardioverter Defibrillator} (ICD), in the patient's body.
Model-based verification can supply rigorous proofs of safety and efficacy. 
In this paper, we build a hybrid system model of the human heart+ICD closed loop, and show it to be a \yhl{STORMED system, a class of o-minimal hybrid systems that admit finite bisimulations.}
In general, it may not be possible to compute the bisimulation.
We show that approximate reachability can yield a finite \emph{simulation} for STORMED systems, which improves on the existing verification procedure.
In the process, we show that certain compositions respect the STORMED property.
Thus it is possible to model check important formal properties of ICDs in a closed loop with the heart, such as delayed therapy, missed therapy, or inappropriately administered therapy. 
The results of this paper are theoretical \yhl{and motivate the creation of concrete model checking procedures for STORMED systems.}
\end{abstract}

\section{Introduction}
\label{sec:intro}
\begin{figure}[t]
	\centering
	\includegraphics[scale=0.3]{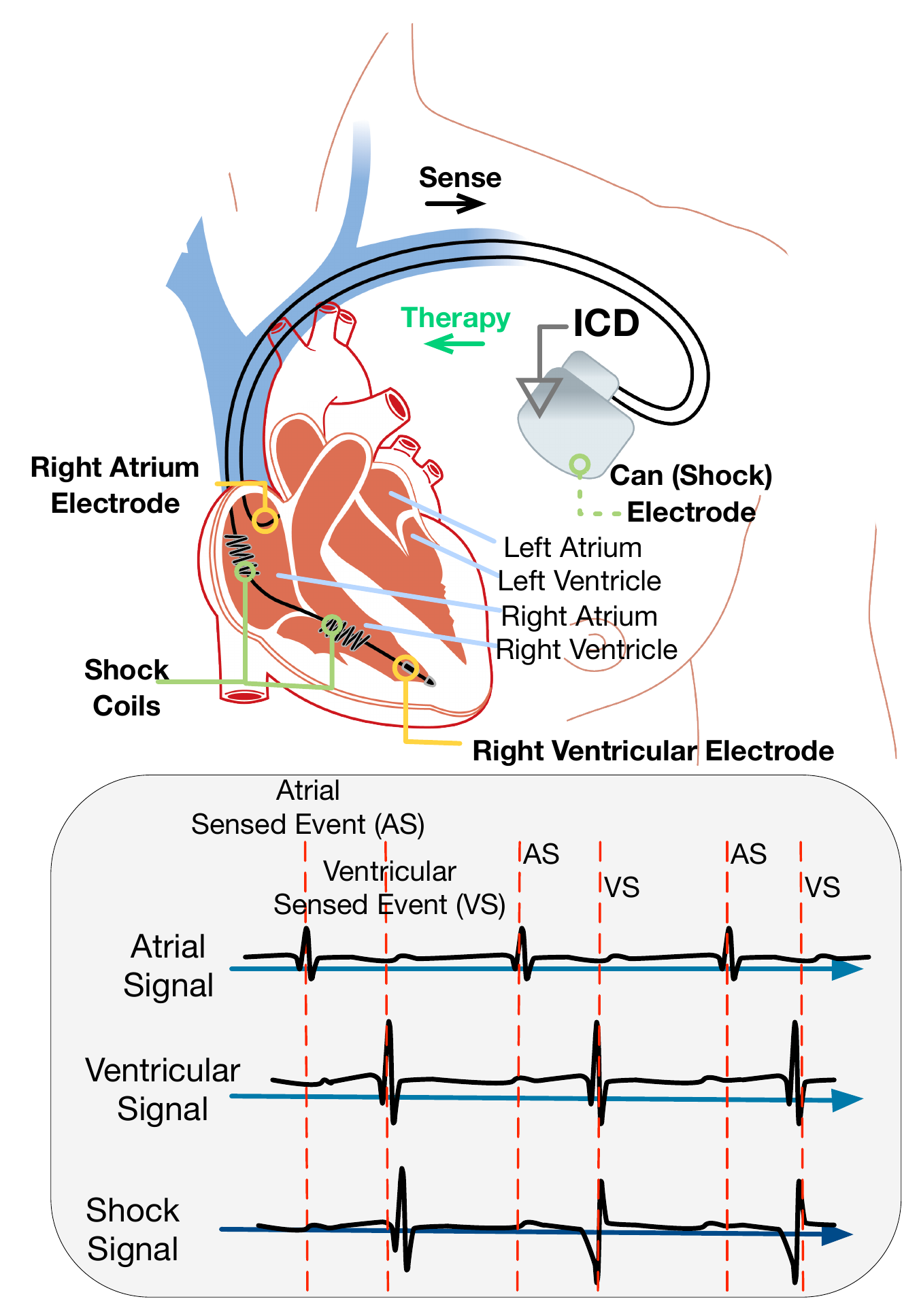}
	\vspace{-10pt}
	\caption{\small ICD connected to a human heart via two electrodes. The ICD monitors three electrical signals (known as electrograms) traversing the heart muscle.}
	\label{fig:icd}
	\vspace{-10pt}
\end{figure}

\acp{ICD} are life-saving medical devices.
An \ac{ICD} is implanted under the shoulder, and connects directly to the heart muscle though two electrodes and continuously measures the heart's rhythm (Fig. \ref{fig:icd}).
If it detects a potentially fatal accelerated rhythm known as Ventricular Tachycardia (VT), the \ac{ICD} delivers a high-energy electric shock or sequence of pulses through the electrodes to reset the heart's electrical activity.
Without this therapy, the VT can be fatal within seconds of onset.
In the US alone, 10,000 people receive an \ac{ICD} every month.
Studies have presented evidence that patients implanted with \acp{ICD} have a mortality rate reduced by up to 31\% \cite{maditrit}.

Unfortunately, \acp{ICD} suffer from a high rate of \emph{inappropriate therapy} due to poor detection of the current rhythm on the part of the \ac{ICD}.
In particular, a class of rhythms known as SupraVentricular Tachycardias (SVTs) can fool the detection algorithms.
Inappropriate shocks increase patient stress, reduce their quality of life, and are linked to increased morbidity \cite{shock_mortality}.
Depending on the particular ICD and its settings, the rates of inappropriate therapy can range from 46\% to 62\% of all delivered therapy episodes \cite{GoldABBTB11_RIGHTresults}.
Current practice for \ac{ICD} verification relies heavily on testing and software cycle reviews.
With the advent of computer models of the human heart, \emph{Model-Based Design} (MBD) can supply rigorous evidence of safety and efficacy. 
This paper presents hybrid system models of the human heart and of the common modules of \acp{ICD} currently on the market, and shows that the closed loop formed by these models is \emph{formally verifiable}.
The objective is to develop model checkers for \acp{ICD} to further their MBD process.

\begin{figure*}[t]
	\centering
	\vspace{-10pt}
	\includegraphics[scale=0.28]{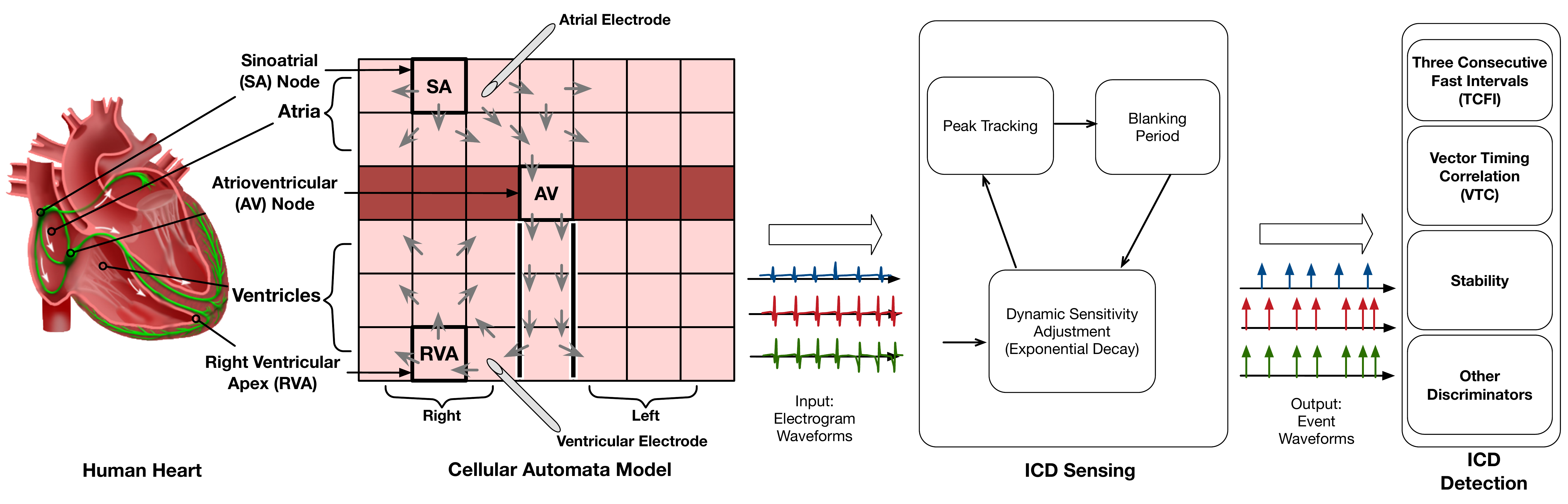}
	\vspace{-10pt}
	\caption{\small The whole heart is modeled as a 2D mesh of cells (Section \ref{sec:heartcellularautomata}). The \ac{ICD} electrodes are shown in the right atrium and ventricle. The electrogram signals measured through the electrodes are processed by the sensing module (ICD Sensing, see Section \ref{sec:sensing}). The detection algorithm (Section \ref{sec:discriminators}) determines the current rhythm using the processed signal (ICD Detection).}
	\label{fig:overview}
	\vspace{-10pt}
\end{figure*}
No work exists on \ac{ICD} verification. 
Earlier work on verification of medical devices (formal or otherwise) focuses on pacemakers.
In \cite{TACAS12} the authors developed timed automata models of the whole heart+pacemaker loop which allows verification of LTL properties.
In \cite{Chen14_Quantitative} the authors perform probabilistic testing of Hybrid I/O automata models of heart and pacemaker.
However, they can not be symbolically verified.
Later work on pacemakers \cite{Mery} develops a formalized \ac{CA} model of the heart and uses Event-B for expressing its properties, and in \cite{HuangFMFK14_VerifCardiacCells} invariants of pacemaker and cardiac cells are verified.
The \ac{ICD} algorithms are more complex than a pacemaker's: an \ac{ICD} measures the timing of events, but also measures and processes the \emph{morphology} of the electrical signal in the heart to distinguish many types of arrhythmias.
Thus, we need three models for \ac{ICD} verification: a timing and voltage model of the heart, a model of the \ac{ICD}'s algorithms, and a model for voltage measurement by the \ac{ICD} electrodes.
This takes the model out of the realm of timed automata and into hybrid automata proper.
\yhl{More generally, approaches to approximate verification of similar hybrid systems include falsification of general Metric Temporal Logic properties \cite{CameronFMS15_ArtificialPancreas} and $\delta$-reachability~\cite{KongGCC15_dreach}.}

The first contribution of this paper is to develop a hybrid system model of the heart, the \ac{ICD} measurement process, and of the algorithmic components of \acp{ICD} from most major manufacturers on the market \yhl{(Fig.~\ref{fig:overview})}.
We show that the composition of these three models admits a finite bisimulation \cite{AlurHLP00ieee}.
The \ac{ICD} models presented here are the first formalization of \ac{ICD} operation to the best of our knowledge.

To establish this result we use the theory of STORMED hybrid systems \cite{VladimerouPVD08_STORMED}, a class of hybrid systems that have finite bisimulations.
Our second contribution is two general results for STORMED systems.
First we prove that parallel compositions of STORMED systems yield STORMED systems.
Secondly, we show that any definable over-approximate reach tubes can replace the exact trajectories of a STORMED system, yielding a system that still admits a finite simulation (but no longer a bisimulation). 
Finally, we show that the reach sets computed by the reachability tool SpaceEx \cite{FrehseCAV11} \yhl{(a widely used and scalable reachability tool) are definable and so can be used to build the simulation.
Thus SpaceEx can be used as part of a model checker for STORMED systems.}

Our interest in not simply in a particular manufacturer's arrhythmia detection algorithm: rather, we are interested in those components that are common to most of them, thus making our results relevant to them.
The components we model or some variation on them are included in the \acp{ICD} of Boston Scientific, Medtronic, Saint-Jude Medical and Biotronik.
This is the first example of a practical STORMED system that the authors are aware of.

\textbf{Organization}. 
Section \ref{sec:preliminaries} covers some preliminaries on hybrid systems.
Sections \ref{sec:heartcellularautomata} presents the heart model,
and Sections \ref{sec:sensing}-\ref{sec:discriminators} model the \ac{ICD}.
Sections \ref{sec:compositionality} and \ref{sec:simulationAprox} prove general results on STORMED systems: namely that a definable over-approximation of the flows such as that computed by SpaceEx preserves finiteness of the simulation, and that compositions of STORMED systems are STORMED.

%
%
%
%
%
%
%
%

\section{Hybrid systems and simulations}
\label{sec:preliminaries}

This section presents fairly standard definitions on 
hybrid systems and their simulations \cite{AlurHLP00ieee}.
It also defines STORMED hybrid systems, which admit finite bisimulations \cite{VladimerouPVD08_STORMED}.

\subsection{Transition and hybrid systems}
\label{sec:transition systems}

%
\begin{defn}
	\label{defn:transition system}
	A \emph{transition system} $T = (Q,\labelSet,\trans{},Q_0)$ consists of a set of states $Q$, a set of events $\labelSet$ , a transition relation $\trans{} \subset Q \times \labelSet \times Q$, a set of initial states $Q_0$. 
	We write $q \trans{\slabel}q'$ to denote a transition element $(q,\slabel,q') \in \trans{}$.
	Given $P\subset Q$, we define $Post_\slabel(P) \defeq \{q'\;|\;\exists q\in P. q \trans{\slabel}q'\}$
	Given an equivalence relation $\sim$ on $Q$, the \emph{quotient system} $T/\sim$ is
	$T/\sim = (Q/\sim, \{*\}, \trans{}_\sim, Q_0/\sim)$
	where $[q] \trans{*}_\sim [q']$ iff $q \trans{\slabel} q'$ for some $\slabel \in \labelSet$.
	Here $[q]$ is the equivalence class of $q$ and $Q/\sim$ is the set of equivalence classes of $\sim$.
\end{defn}

\begin{defn}
	\label{defn:simulation}	
	Given two transition systems $T_1$ and $T_2$ with the same state space $Q$,
	a \emph{simulation} relation from $T_1$ to $T_2$ is a relation $\simu \subset Q \times Q$ such that 
	for all $(q_1,q_2) \in \simu$, if $q_1 \trans{\slabel}_1 q_1'$, there exists a $q_2' \in Q$ s.t. $q_2 \trans{\slabel}_2 q_2'$ and $(q_1',q_2') \in \simu$.
	A \emph{bisimulation relation} between $T_1$ and $T_2$ is both a simulation relation from $T_1$ to $T_2$ and from $T_2$ to $T_1$.
\end{defn}
%
The bisimulation $\bisimu$ is said to \emph{respect} $\sim$ if $(q,q') \in \bisimu \implies q \sim q'$.
The following algorithm, if it terminates, yields a finite bisimulation for $T$ that respects the given equivalence relation~\cite{AlurHLP00ieee}.
Moreover, it is the \emph{coarsest} bisimulation (with respect to inclusion) that respects $\sim$.
\begin{algorithm}[t]
		\caption{Computing a bismimulation respecting $\sim$}
		\label{algo:bisimulation}
		\begin{algorithmic}
			\Require Transition system $T = (Q,\labelSet,\trans{},Q_0)$, equivalence relation $\sim$.
			\State Set $\simu = Q/\sim$			
			\While{$\exists P,P' \in \simu$ and $\slabel \in \labelSet$ s.t. $\emptyset \neq P' \cap Post_\slabel(P) \neq P'$}
				\State Set $\simu = \simu \setminus \{P'\} \cup \{P' \cap Post_\slabel(P) , P' \setminus Post_\slabel(P) \}$
			\EndWhile	
			\State Return $\simu$
		\end{algorithmic}
\end{algorithm}
Given a set of atomic propositions $AP$, if $\sim$ is s.t. $q \sim q'$ iff both states satisfy exactly the same set of atomic propositions, then model checking temporal logic properties can be done on the finite bisimulation instead of the possibly infinite $T$.


%
\begin{defn}
	\label{defn:hybrid system}	
	A \emph{hybrid automaton} is a tuple \[\Sys = (\stSet,\modeSet,\hsSet_0,\{f_\mode\}, Inv,E, \{\reset_{ij}\}_{(i,j)\in E}, \{\guard_{ij}\}_{(i,j)\in E})\] where 
		 $\stSet \subset \Re^n$ is the continuous state space equipped with the Euclidian norm $\|\cdot\|$, 
		$\modeSet \subset \Ne$ is a finite set of modes,
		 $\hsSet_0 \subset \stSet \times \modeSet$ is an initial set,
		 $\{f_\mode\}_{\mode \in \modeSet}$ determine the continuous evolutions with unique solutions,
		 $Inv: \modeSet \rightarrow 2^\stSet$ defines the invariants for every mode,
		 $E \subset \modeSet^2$ is a set of discrete transitions,
		 \yhl{$\guard_{ij} \subset \stSet$ is guard set for the transitions (so $\Sys$ transitions $i \rightarrow j$ when $\stPt \in \guard_{ij}$),
		 $\reset_{ij}: \stSet \rightarrow \stSet$ is an edge-specific reset function.}
		 \\
		 Set $\hsSet = \modeSet\times \stSet$.
		 Given $(\mode,\stPt_0) \in \hsSet$, the \emph{flow} $\theta_{\mode}(;\stPt_0):\Re_+ \rightarrow \Re^n$ is the solution to the IVP $\dot{x}(t) = f_\mode (x(t))$, $\stPt(0)=\stPt_0$.
\end{defn}
The associated transition system is $T_\Sys = (\hsSet,  E \cup \{\tau\},\trans{},\hsSet_0)$ 
with $\trans{} = (\bigcup_{e \in E} \trans{e}) \cup \trans{\tau}$ 
where $(i,\stPt) \trans{e} (j,y)$ iff $e = (i,j), \stPt \in \guard_{ij}, y = \reset_{ij}(\stPt)$ and $(i,\stPt) \trans{\tau} (j,y)$ iff $i = j$ and there exists 
a flow $\theta_i(\cdot;x)$ of $\Sys$ and $t\geq 0$ s.t. $\theta_i(t;x)=y$ and $\forall t' \leq t$, $\theta_i(t';x) \in Inv(i)$.
For a set $P \subset \hsSet$,$P_{|\stSet}$ denotes its projection onto $\stSet$, 
and $P_{|\modeSet}$ its projection onto $\modeSet$. 
\begin{defn}
	\label{defn:reachability operators}[Reachability]
	Let $\Sys$ be a hybrid system with hybrid state space $\hsSet$, 	 
	$I = [0,b) \subset [0,+\infty)$ be a (possibly unbounded) interval, 
	$t \in I$, 
	and $\epsilon >0$.
	The \emph{$\epsilon$-approximate continuous reachability operator}, 
	$\Rc^{\epsilon}_t : 2^\hsSet \rightarrow 2^\hsSet$ is given by
	\begin{eqnarray*}
		\Rc^{\epsilon}_t(P) = \{(i,\stPt) \in \stSet | \exists x_0 \in P_{|\stSet}, t \geq 0. 
		||\theta_i(t;x_0) - \stPt|| \leq \epsilon\} 
	\end{eqnarray*}
	where $P = \{i\}\times W$, $W \subset Inv(i)$.
	Define also $\Rc^{\epsilon}_I(P) = \cup_{t\in I} \Rc^{\epsilon}_t(P)$.
	The (exact) \emph{discrete reachability operator} is:
	\begin{eqnarray*}
	\Rc_{d}(P) &=& \cup_{j: (i,j) \in E} \reset_{ij}(P \cap G_{ij})
	\end{eqnarray*}
\end{defn}
For a hybrid system, $Post_\slabel$ computes the forward reach sets, and is implemented by $\Rc^0_{[0,\infty)}$ and $\Rc_d$. 
Algorithm \ref{algo:bisimulation}, applied to $T_\Sys$, implements the following iteration, 
in which 
$\Fc_t(\partition)$ is the coarsest bisimulation with respect to $\trans{\tau}$\footnote{I.e., $\Ft$ only considers the continuous transition relation. Namely, it is a bisimulation of $T_\Sys^c \defeq (Q/\sim,\{*\},\trans{\tau},Q_0/\sim)$.} 
respecting the partition $\partition$, 
and 
$\Fc_d(\partition) \defeq \{(h_1,h_2)  \;|\; (h_1 \trans{e} h_1') \implies (\exists e' \in E, h_2' \; . h_2 \trans{e'} h_2' \land h_1' \equiv_\partition h_2') \} \cap \partition$ \cite{VladimerouPVD08_STORMED}:
\begin{eqnarray}
W_0 = \Ft(Q/\sim), \quad\forall i\geq 0,\; W_{i+1} = \Ft(\Fd(W_i))
\end{eqnarray}
This iteration (equivalently, Alg.~\ref{algo:bisimulation}) does not necessarily terminate for hybrid systems because the reach set might intersect a given block of $Q/\sim$ an infinite number of times (see \cite{LaFerrierePS00_Ominimal} for an example).
The class of systems introduced in the next section has the property that Algorithm \ref{algo:bisimulation} does terminate for it and returns a finite $\simu$.



\subsection{O-minimality and STORMED systems}
\label{sec:ominimality}
We give a very brief introduction to o-minimal structures.
A more detailed introduction can be found in \cite{LaFerrierePS00_Ominimal} and references therein.
We are interested in sets and functions in $\Re^n$ that enjoy certain finiteness properties, called order-minimal sets (o-minimal).
These are defined inside \emph{structures} $\Ac = (\Re,<, +,-,\cdot,\exp,\ldots)$.
The subsets $Y \subset \Re^n$ we are interested in are those that are \emph{definable} using first-order formulas $\formula$: $Y = \{(a_1,\ldots,a_n) \in \Re^n \;|\;  \formula(a_1,\ldots,a_n)\}$.
(First-order formulas use the boolean connectives and the quantifiers $\exists,\forall$).
The atomic propositions from which the formulas are recursively built allow only the operations of the structure $\Ac$ on the real variables and constants, and the relations of $\Ac$ and equality.
For example $2x-3.6y < 3z$ and $x=y$ are valid atomic propositions of the structure $\Lc_\Re=(\Re,<, +,-,\cdot)$, while $cosh(x) < 3z$ is not because $cosh$ is not in the structure.
These structures are already sufficient to describe a set of dynamics rich enough for our purposes and for various classes of linear systems.
\begin{defn}
	\label{defn:ominimal struct}	
	A theory of $(\Re,\ldots)$ is \emph{o-minimal} if the only definable subsets of $\Re$ are finite unions of points and (possibly unbounded) intervals.	
	A function $f:x \mapsto f(x)$ is o-minimal if its graph $\{(x,y) \;|\; y=f(x)\}$ is a definable set.
\end{defn}
We use the terms o-minimal and definable interchangeably, and they refer to $\Lc_{\exp}= (\Re,<, +,-,\cdot,\exp)$ which is known to be o-minimal.
The dot product between $x,y\in \Re^n$ is denoted $x \cdot y$, and $d(Y,S)=\inf \{ \|y-s\| \;|\; (y,s) \in Y\times S \}$.
\begin{defn}\cite{VladimerouPVD08_STORMED}.
	\label{defn:stormed system}	
	A \emph{STORMED hybrid system} (SHS) $\SHS$ is a tuple $(\Sys,\Ac, \phi,b_-,b_+, d_{min}, \epsilon,\zeta)$ where $\Sys$ is a hybrid automaton, $\Ac$ is an o-minimal structure, $d_{min}, \epsilon, \zeta$ are positive reals, $b_-,b_+ \in \Re$ and $\phi \in \stSet$ such that:
	\\
	\textbf{(S)} The system is $d_{min}$-separable, meaning that for any $e=(\mode, \mode ')\in E$ and $\mode ''\neq \mode'$,$d(\reset_e(\guard_{(\mode ,\mode ')}), \guard_{(\mode' ,\mode '')})>d_{min}$
	\footnote{\yhl{The original definition of separability \cite{VladimerouPVD08_STORMED} required the guards themselves to be separated, which is insufficient to guarantee that if $\Sys$ flows, it flows a uniform minimum distance along $\phi$. Indeed assume the guards are separated. If $x \in \guard_{(\mode,\mode')}$ and $y = \reset_{(\mode,\mode')}(x)$, it can be that $y \in \guard_{(\mode',\mode'')}$ and thus a jump happens, even though $\guard_{(\mode,\mode')}$ and $\guard_{(\mode',\mode'')}$ are separated. Therefore we need $d(y,\guard_{\mode',\mode''}) > d_{min}$ for all $y \in \reset_e(\guard_e)$, which is the condition we use in Def.~\ref{defn:stormed system}. The properties of SHS, in particular the existence of finite bisimulation, are therefore preserved by this change.}}
	\\
	\textbf{(T)} The flows (i.e., the solutions of the ODEs) are Time-Independent with the Semi-Group property (TISG), meaning that for any $\mode \in \modeSet, \stPt \in \stSet$, the flow $\theta_\mode$ starting at $(\mode , x)$ satisfies: 1) $\theta_{\mode}(0;x) = x$, 2) for every $t,t' \geq 0$, $\theta_{\mode}(t+t';x) = \theta_{\mode}(t'; \theta_\mode(t;x))$
	\\
	\textbf{(O)} All the sets and functions of $\Sys$ are definable in the o-minimal structure $\Ac$
	\\
	\textbf{(RM)} The resets and flows are monotonic with respect to the same vector $\phi$, meaning that \\
	1) (Flow monotonicity) for all $\mode \in \modeSet$, $x \in \stSet$ and $t,\tau \geq 0$, $\phi \cdot (\theta_\mode (t+\tau;x) - \theta_\mode (t;x) ) \geq \epsilon ||\theta_\mode (t+\tau;x) - \theta_\mode (t;x) ||$, 
	and \\
	2) (Reset monotonicity) for any edge $(\mode,\mode') \in E$ and any $x^-,x^+ \in \stSet$ s.t. $x^+ = R_{\mode ,\mode'}(x^-)$, 
	\begin{compactenum}
		\item if $\mode = \mode'$, then either $x^-=x^+$ or $\phi \cdot (x^+-x^-)\geq \zeta$
		\item if $\mode \neq \mode'$, then $\phi\cdot (x^+-x^-) \geq \epsilon ||x^+-x^-||$
	\end{compactenum}

	\textbf{(ED)} Ends are Delimited: for all $e \in E$ we have $\phi \cdot x \in (b_- , b_+)$ for all $x \in G_{e}$
\end{defn}
Intuitively, the above conditions imply the trajectories of the system always move a minimum distance along $\phi$ whether flowing or jumping, which guarantees that no area of the state space will be visited infinitely often. 
This is at the root of the finiteness properties of STORMED systems.
The following result justifies the interest in STORMED systems: they admit finite bisimulations.
\begin{thm}\cite{VladimerouPVD08_STORMED}
	\label{thm:stormed finite bisimu}	
	Let $\Sys$ be a STORMED hybrid system, and let $\partition$ be an o-minimal partition of its hybrid state space. 
	Then $\Sys$ admits a finite bisimulation that respects $\partition$.
\end{thm}
We need the following result in what follows.
\begin{prop}
	\label{prop:ED}
	If the state space $\stSet$ of a hybrid automaton $\Sys$ is bounded, then its guards have delimited ends.
\end{prop}
\begin{prf}
	For all guard sets $G$ and all $x \in G$, $||\phi \cdot x || \leq ||\phi|| \cdot ||x|| \leq ||\phi||.\max\{||x||, x\in \stSet \} < \infty$.
\end{prf}

\section{Heart model}
\label{sec:heartcellularautomata}
For the verification of \acp{ICD},
we adopt the \acf{CA}-based heart model developed in \cite{Spector11_Emergence},\cite{CorreaEtAl11_EGMFractionation}.
This model lies in-between high spatial fidelity but slow to compute PDE-based whole heart models  \cite{vfiborganization_Tusscher07}, and low spatial fidelity but very fast-to-compute automata-based models \cite{TECS}.
PDE-based models are not currently amenable to formal verification, both theoretically and practically.
\yhl{Models based on ionic currents \cite{Islam1MGSG14_CompositionalityCells} might be more accurate but are likely to be more computationally expensive.}
Timed automata models can not simulate the electrograms needed for \ac{ICD} verification.
\ac{CA}-based models are appealing due to their intuitive correspondence with the heart's anatomy and function and their relative computational simplicity.
\ac{CA}-based models were used in \cite{Mery},\cite{BartocciCBESG09_HIOAmodeling} and \cite{Chen14_Quantitative}.
This paper's model also has the important advantage of forming the basis of software used to train electrophysiologists, and allows interactive simulation of surgical procedures like ablation \cite{visibleep}.
\yhl{In particular, it can simulate fibrillation and other tachycardias.}%

\begin{figure}[t]
\centering
\includegraphics[scale=0.4]{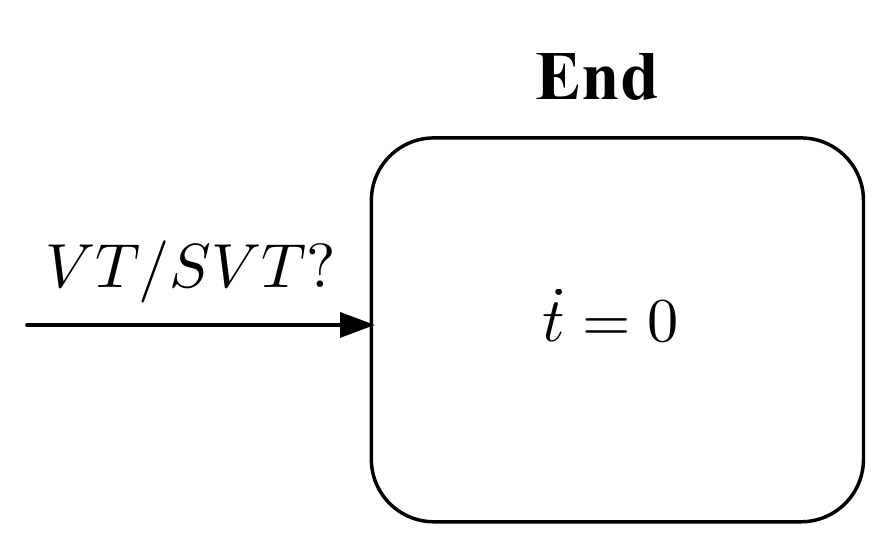}
\caption{When the ICD makes a VT/SVT decision, all systems transition to mode End.}
\label{fig:endMode}
\end{figure}

\textbf{This paper's automata:}All hybrid automata in this paper have the whole state space as invariants and transitions are urgent (taken immediately when the guard is enabled).
We also observe that, as will be seen in Section \ref{sec:discriminators},
i) the \ac{ICD} will always reach a decision of VT or SVT in finite time, 
ii) at which point it resets its \yhl{controlled (software) variables} so new values are computed for the next arrhythmia episode.
So while the heart can beat indefinitely, for the purposes of \ac{ICD} verification, 
there's a uniform upper bound on the length of time of any execution.
Let $D \geq 0$ be this duration ($D$ is on the order of 30sec depending on device settings).
Also, the \ac{EGM} voltage signal $\egm$ has upper and lower bounds $\overline{\egm}$ and $\underline{\egm}$.
\yhl{Therefore, every mode of every automaton in what follows has a transition to mode End shown in Fig.~\ref{fig:endMode}.
We don't show these transitions in the automata figures to avoid congestion.
}

\subsection{Cellular automata model}
The heart has two upper chambers called the \emph{atria} and two lower chambers called the \emph{ventricles} (Fig. \ref{fig:icd})
The synchronized contractions of the heart are driven by electrical activity.
Under normal conditions, the SinoAtrial (SA) node (a tissue in the right atrium) spontaneously \emph{depolarizes}, producing an electrical wave that propagates to the atria and then down to the ventricles (Fig.\ref{fig:overview})
In this model, the myocardium (heart's muscle) is treated as a 2D surface (so it has no depth), and discretized into \emph{cells}, which are simply regions of the myocardium (Fig. \ref{fig:overview}). 
Thus we end up with $N^2$ cells in a square $N$-by-$N$ grid.
A cell's voltage changes in reaction to current flow from neighboring cells, and in response to its own ion movements across the cell membrane.
This results in an \emph{\ac{AP}}.

\begin{figure}[t]
	\centering
	\includegraphics[scale=0.26]{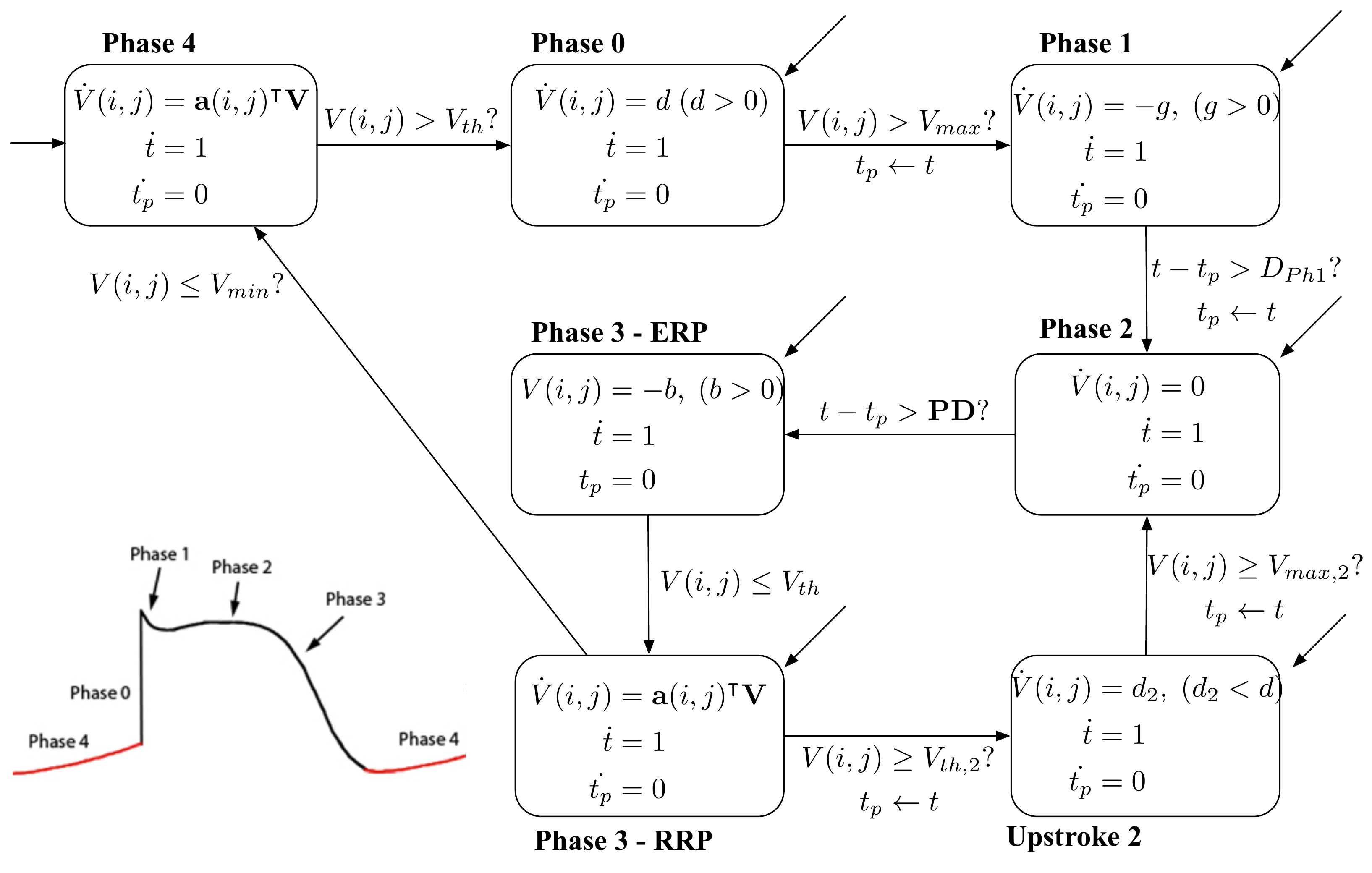}
	\vspace{-10pt}
	\caption{Hybrid model $\Sys_c$ of one cell of the heart model. AP figure from \cite{eplab}. 
		$V_{th,2}>V_{th}$, $V_{max,2}<V_{max}$}
	\vspace{-20pt}
	\label{fig:cellaut}
\end{figure}

Fig. \ref{fig:cellaut} shows how the \ac{AP} is generated by a given cell \cite{Klabunde_CVEPconcepts}:
in its quiescent mode (Phase 4), a cell $(i,j)$ in the grid has a cross-membrane voltage $V(i,j,t)$ equal to $V_{min} < 0$.
As it gathers charge, $V(i,j,t)$  increases until it exceeds a threshold voltage $V_{th}$.
\yhl{In Phase 0}, the voltage then experiences a very fast increase (Phase 0), called the upstroke, to a level $V_{max} > 0$, after which it decreases \yhl{(Phase 1)} to a plateau \yhl{(Phase 2)}.
It stays at the plateau level for a certain amount of time \textbf{PD} then decreases linearly to below $V_{th}$ (Phase 3 - ERP).
Once below $V_{th}$ it is said to be in the Relative Refractory Period (Phase 3 - RRP) .
\yhl{In Phase 3 - RRP}, the cell can be depolarized a second time, albeit at a higher threshold $V_{th,2}$, slower and to a lower plateau level $V_{max,2} < V_{max}$ \yhl{(Upstroke 2)}.
Otherwise, when the voltage reaches $V_{min}$ again, the cell enters the quiescent stage again. 
This model is suitable for both pacemaker and non-pacemaker cells, the main differences being in the duration of the plateau (virtually non-existent for pacemaker cells), and the duration of phases 0 and 4 (both are shorter for pacemaker cells).

In Fig. \ref{fig:cellaut}, $V(i,j) \in \Re$ denotes the voltage in cell $(i,j)$ of the grid, and \yhl{$V =(V(1,1),\ldots, V(N^2,N^2))^T$} in $\Re^{N^2}$ groups the cross-membrane voltages of all cells in the heart.
The whole heart model $\Sys_{CA}$ is the parallel composition of these $N^2$ single-cell models. 
The $(i,j)^{th}$ cell's voltage at time $t$ in Phase 4 depends on that of its neighbors and its own as follows \cite{Spector11_Emergence}
\begin{eqnarray}
\dot{V}(i,j,t) &=& \frac{1}{R_h}[V(i-1,j,t)+V(i+1,j,t) - 2V(i,j,t)] 
\nonumber \\ 
&& +  \frac{1}{R_v}[V(i,j-1,t) +  V(i,j+1,t) - 2V(i,j,t)]  
\nonumber\\
&=& a(i,j)^TV(t), \; a(i,j) \in \Re^{N^2}
\;
\end{eqnarray}
where $R_h$, $R_v$ are conduction constants that can vary across the myocardium.
Thus $V$ evolves according to a linear ODE $\dot{V} = AV$ where $A$ is the matrix whose rows are the $a(i,j)$. 
The two states $t$ and $t_p$ are clocks.
Clock $t_{p}$ keeps track of the value of the last discrete jump. 
We will use this arrangement in all our models: it avoids resetting the clocks which preserves Reset Monotonicity.

\acp{ICD} observe the electrical activity through three channels (Fig.~\ref{fig:icd}).
Each signal is called an \acf{EGM} signal.
The signal read on a channel is given by \cite{CorreaEtAl11_EGMFractionation}:
\begin{equation}
	\label{eq:vegm}
	\egm(t) = \frac{1}{K} \sum_{i,j} \left(\frac{1}{||p_{i,j}-p_0|| } - \frac{1}{||p_{i,j}-p_1||}\right) \dot{V}(i,j,t)
\end{equation}
\yhl{where $\|\cdot\|$ is the Euclidian norm, $p_0$ and $p_1$ are the electrodes' positions and $p_{i,j}$ is the position of the $(i,j)^{th}$ cell on the 2D myocardium ($p_0,p_1,p_{i,j} \in \Re^2$). 
Positions $p_0,p_1$ should be chosen different from $p_{i,j}$ to avoid infinities.}

\textbf{Extensions}. 
The Action Potential Duration (APD) restitution mechanism of heart cells as modeled in \cite{Spector11_Emergence} can be included in this model without changing its formal properties.
\yhl{More detailed APD restitution models exist~\cite{Grosu09_LearningEmergent}.
Also, note that cell topology (the way cells are connected to each other) is not a factor in determining the STORMED property, so other topologies than a rectangular mesh may be used.}

We now state and prove the main result of this section.
\begin{thm}
	Let $\Sys_{CA}$ be the whole heart cellular automaton model obtained by parallel composition of $N^2$ models $\Sys_c$ with state vector $x = [V, t,t_p,\egm ] \in \Re^{N^2}\times \Re^{3}$.
	Assume that all executions of the system have a duration of $D\geq 0$.
	Then $\Sys_{CA}$ is STORMED.
\end{thm}
\begin{prf}
	We verify each property of STORMED.
In this and all the proofs that follow, the approach is the same: $(ED)$ holds by Prop.~\ref{prop:ED} because our state spaces are bounded.
After establishing properties $(S), (T)$ and $(O)$, we draw up the constraints on $\phi$ and $\varepsilon$ imposed by reset and flow monotonicity (property (RM)). 
Then we argue that these constraints can be solved for $\phi$ and $\varepsilon$.
Often there is more than one solution and we just point to one.

\textbf{(S)} Separability holds because $V_{min} < V_{th}< V_{th,2} < V_{max,2} < V_{max}$ and $PD>0,D_{Ph_1}>0$. 
For example, on transition \textbf{Phase 4} $\rightarrow$ \textbf{Phase 0}, $V(i,j)=V_{th}$, which is separated from the next guard $\{V(i,j) > V_{max}\}$ by $|V_{max}-V_{th}|$.
\\
\textbf{(T)} All flows are linear or exponential and thus are TISG.
\\
\textbf{(O)} The flows, resets and guard sets are all definable in $\Lc_{\exp}$.
In particular the flow of $\dot{V} = AV$ is exponential with real exponent, and $\egm$ is a sum of exponentials and linear terms.
\\
\textbf{(RM)}
We seek a vector $\phi = (\phi_V,\phi_t,\phi_p,\phi_\egm)^T \in \Re^{N^2+3}$ such that resets and flows are monotonic along $\phi$.
Only transitions $p \rightarrow q \neq p$ are to be found in $\Sys_{CA}$, during which only $t_p$ is reset. 
Always, $t_p^+ = t \geq t_p^-$, thus the reset is indeed monotonic as can be seen by choosing any $\varepsilon >0$ and $\phi_p > \varepsilon$.

Monotonic flows: $\phi$ must also be such that in all modes:
\begin{equation*}
\phi \cdot (\theta_\ell(t+\tau;x) - \theta_\ell(t;x)) \geq \varepsilon ||\theta_\ell(t+\tau;x) - \theta_\ell(t;x)||
\end{equation*}
Decomposing, we want 
\begin{eqnarray}
\label{eq:monotonic flow ca}
&&\phi_V \cdot(V(t+\tau) - V(t)) + \phi_t \tau + \phi_p \cdot 0 
\\
&&\quad+ \phi_\egm \cdot (\egm(x,t+\tau) - \egm(x,t)) \geq \varepsilon ||\theta_\ell(x,t+\tau) - \theta_\ell(x,t)||
\nonumber 
\end{eqnarray}
Now note that all flows have bounded derivatives in every bounded duration of flow and are thus Lipschitz. 
Let $L_V$ be the Lipshitz constant of $V(t)$ and $L_{\egm}$ that of $\egm(t)$.
Then on the LHS of the above inequality we have 
$\phi_V \cdot(V(t+\tau) - V(t)) + \phi_\egm \cdot(\egm(t+\tau) - \egm(t)) \geq -\phi_V L_V \tau  -\phi_\egm L_{\egm} \tau$.
On the RHS we have  
$\varepsilon (L_V\tau + L_{\egm}\tau + \tau) \geq \varepsilon ( ||V(t+\tau) - V(t)|| + ||\egm(t+\tau) - \egm(t) ||  + \tau) \geq \varepsilon ( ||\theta_\ell(x,t+\tau) - \theta_\ell(x,t)||)$
Thus \eqref{eq:monotonic flow ca} is satisfied if the stronger inequality 
\[-\phi_V L_V \tau  -\phi_\egm L_{\egm} \tau + \phi_t \tau \geq \varepsilon (L_V\tau + L_{\egm}\tau + \tau) \]
is satisfied.
But this can be achieved by, for example, choosing $\phi_V = \phi_\egm = 0$ and $\phi_t \geq \varepsilon(L_V+L_{\egm}+1)$.
\\
\textbf{(ED)} Our system has bounded state spaces: $V$ and $\egm$ are voltages typically in the range $[-80,60]$ mV and $t_p \leq t \leq D$.
So (ED) holds by Lemma \ref{prop:ED}. 
\end{prf}
\section{ICD Sensing}
\label{sec:sensing}
\begin{figure}[t]
	\centering
	\includegraphics[scale=0.35]{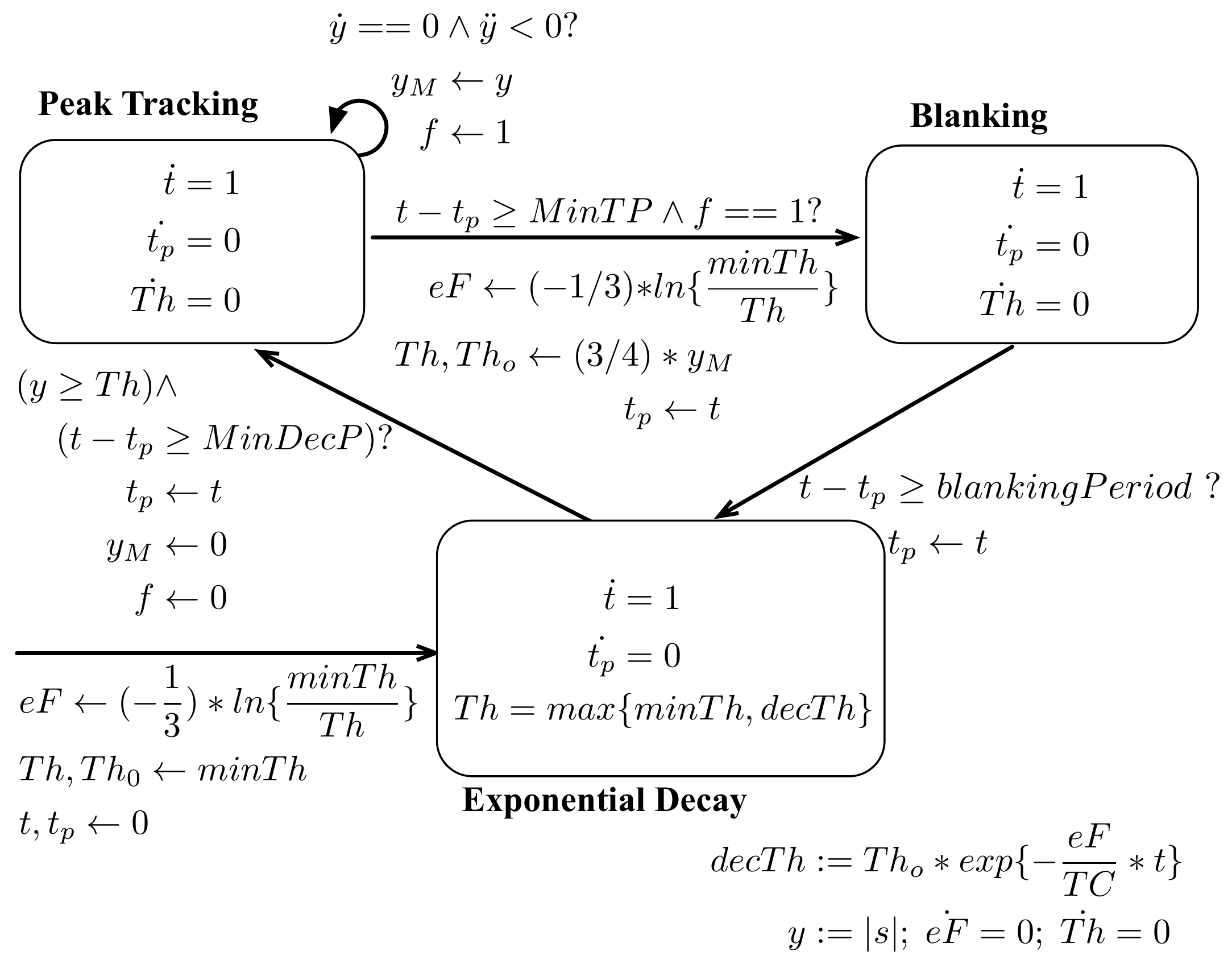}
	\vspace{-10pt}
	\caption{\small $\Sys_{Sense}$. States not shown in a mode have a 0 derivative, e.g., $\dot{eF}=0$ in all modes.}
	\vspace{-10pt}
	\label{fig:sensingModel}
\end{figure}
\begin{figure}[t]
	\centering
	\includegraphics[scale=0.3]{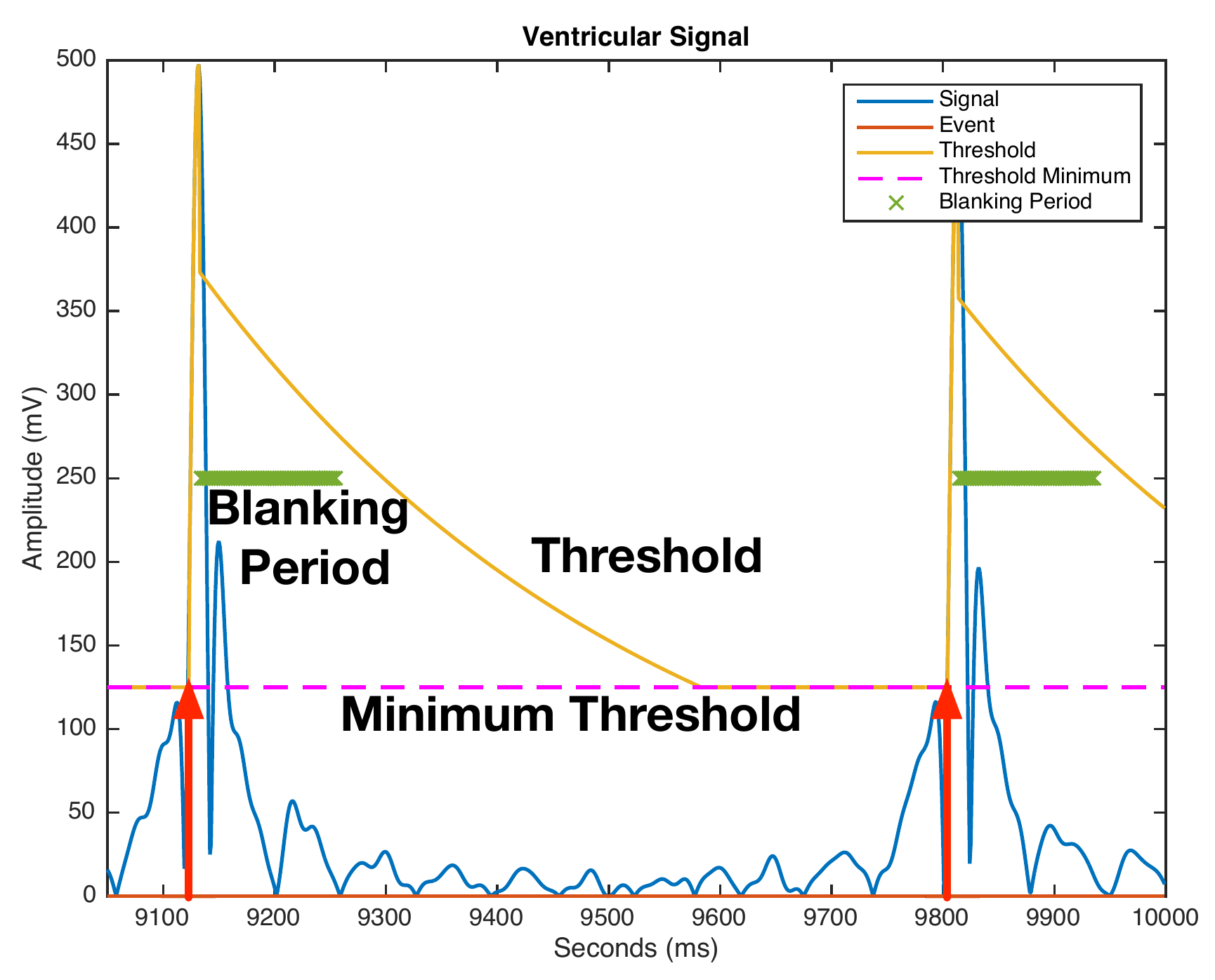}
	\vspace{-10pt}
	\caption{\small Example of dynamic threshold adjustment in ICD sensing algorithm. The shown signal is rectified.}
	\vspace{-10pt}
	\label{fig:sensingExample}
\end{figure}

\emph{Sensing} is the process by which cardiac signals $\egm$ measured through the leads of the \ac{ICD} are converted to cardiac timing events.
The \ac{ICD} sensing algorithm is a threshold-based algorithm which declares events when the signal exceeds a dynamically-adjusted threshold $Th$.

Fig. \ref{fig:sensingModel} shows the model $\Sys_{Sense}$ of the sensing algorithm, and Fig. \ref{fig:sensingExample} illustrates its operation. 
The sensing takes place on the rectified \ac{EGM} signal $y = |\egm|$.
After an event is declared at the current threshold value ($y(t)\geq Th(t)$ in Fig. \ref{fig:sensingModel}), the algorithm tracks the signal in order to measure the next peak's amplitude \yhl{(Peak Tracking)}.
For a duration $MinTP$ (min tracking period) the latest peak is saved in $y_M$.
A variable $f$ indicates that a peak was found.
After a peak is found ($f==1$) and after the end of the tracking period, the algorithm enters a fixed \emph{Blanking Period} \yhl{(Blanking)}, during which additional events are ignored.
\yhl{On the transition to Blanking}, $Th$ and $Th_0$ are set to 3/4 the current value of $y_{M}$ and the exponential factor of decay is updated ($eF=(-1/3)*ln{\frac{minTh}{TH}}$). 
At the end of the blanking period, the algorithm then transitions to the Exponential Decay mode in which $Th$ decays exponentially from $Th_0$ to a minimum level \yhl{(Exponential Decay)}:
$Th(t) = \max(minTh, Th_0\cdot exp(-(eF/TC)t)) $.
The algorithm stays in the Exponential Decay mode for at least a sampling period of $MinDecP$.
Correspondingly, there is a de facto Maximum Decay Period $MaxDecP$ after which the system transitions again to PeakTracking since the signal $y$ is bound to exceed the minimum threshold $minTh$.
Different manufacturers may use a step-wise decay instead of exponential, but the principle is the same.
Local peak detection is modeled via the $\dot{y} = 0 \wedge \ddot{y}<0$ transition.
While $y=|\egm|$ is non-differentiable at 0, the peak will occur away from 0, as shown in Fig. \ref{fig:sensingExample}.
The other states in Fig. \ref{fig:sensingModel} are $t, t_p$ (clocks).
$minTh$ and $TC$ are constant parameters.
\begin{thm}
	\label{thm:sensing}
	$\Sys_{Sense}$ is STORMED.	
\end{thm}
\begin{prf}
	\textbf{(S)} By definition, we only need to consider transitions between different modes to establish separability. 
	For all such transitions, there is a minimum dwell time in the mode before taking the transition, namely $MinTP$ in PeakTracking, $BlankingPeriod$ in Blanking, and  $MinDecP$ in mode ExponentialDecay.
	So the system is separable since there is a uniform minimum flow before jumping.
	\\
	\textbf{(T)} Flows are either constant, (piece-wise) linear, or piece-wise linear and exponential (in the case of $y$ and its derivatives) and therefore are TISG.
	\\
	\textbf{(O)} All the flows, resets and guard sets are definable in $\Lc_{\exp}$.
	(The absolute value and $\max$ functions can be broken down into boolean disjunctions of definable functions, and $t \mapsto \ln(t)$ is o-minimal by o-minimality of $\exp$).
	\\
	\textbf{(RM)} The state is $x = (t, t_p, y, y_M, f, Th, Th_0,eF) \in \Re^8$, and let 
	 $\phi = (\phi_t, \phi_{p}, \phi_y, \phi_m, \phi_f, \phi_{Th},\phi_0,\phi_{eF})$ be the corresponding $\phi$ vector.
	Recall that the \ac{EGM} voltage $\egm$, and so $y=|s|$, is upper-bounded by $V_M$.	
	\\ 
	\textbf{ExponentialDecay $\rightarrow$ PeakTracking}.
	Only $t_p,y_M$ and $f$ are modified, so monotonicity produces the constraint
	 $\phi_p(t-t_p) +\phi_m(0-y_M) + \phi_f(0-1) \stackrel{Want}{\geq} \varepsilon (|t-t_p|+|y_M|+1)$.
	We require the stronger constraint to hold:
	\[\phi_t MinDecP - \phi_m V_M -\phi_f \stackrel{Want}{\geq} \varepsilon(MaxDecP + V_M+1)\]
	\\
	\textbf{PeakTracking $\rightarrow$ PeakTracking}. Only $y_M$ and $f$ are reset. 
	Algebraic manipulation yields $-2V_M\phi_m + \phi_f \stackrel{Want}{\geq} \zeta$
	\\
	\textbf{PeakTracking $\rightarrow$ Blanking}.
	$t_p,eF,Th$ and $Th_0$ are reset, so we get
	\begin{eqnarray*}
	&&\;\phi_p(t-t_p) + \phi_{eF}(-(1/3)\ln(minTh/Th)-eF) 
	\\
	&&+\phi_{Th}(3y_M/4-Th) +\phi_0(3y_M/4-Th_0)
	\\
	&&\geq \varepsilon(|t-t_p|+ |-\frac{1}{3}\ln(\frac{minTh}{Th})-eF|
	\\
	&&+|\frac{3y_M}{4}-Th|+|\frac{3y_M}{4}-Th_0|)
	\end{eqnarray*}
	
	$Th$ is lower-bounded by $minTh$ at all times, and it is naturally upper-bounded by $V_M$ as the threshold should never exceed the largest possible attainable voltage. 
	By the same token, $0\leq eF \leq (1/3)\ln(V_M/minTh)$.
	Then we want the stronger inequality
	\begin{eqnarray*}
	\phi_p MinTP &+& \phi_{eF}(0-(1/3)\ln(V_M/minTh)
	\\
	&+&\phi_{Th}(-V_M) +\phi_0(-V_M)
	\\
	&\geq& \varepsilon(MaxTP+ |\frac{1}{3}\ln(\frac{V_M}{Th})|+|V_M|+|V_M|)
	\end{eqnarray*}
	\\
	\textbf{Blanking $\rightarrow$ ExponentialDecay}. Only $t_p$ is reset and therefore we want, $\phi_p(t-t_p) \geq \varepsilon(|t-t_p|)$, thus the transition yields $\phi_p \geq \varepsilon$.
	
	The above equations can be simultaneously satisfied.
	The simplest thing would be to set all $\phi$ terms that appear above to 0 except for $\phi_t,\phi_p$ which are calculated accordingly.
	
	The flows can be shown to be monotonic along the same $\phi$ and with the same $\varepsilon$.
	For example, in mode ExponentialDecay, only $t,y$ and $Th$ flow.
	Making use of the $V_M$ bound on $y$, we get the constraint
	$\phi_t \tau - 2V_M\phi_y +\phi_{Th}(Th(t+\tau)-Th(t))\geq \varepsilon(\tau+2V_M + |Th(t+\tau)-Th(t)| )$, 
	which yields $\phi_t \geq \varepsilon$, $\phi_y \leq -\varepsilon$ and $\phi_{Th} \geq \varepsilon$. 
	Similarly for the rest.	
\end{prf}

\section{Arrhythmia detection}
\label{sec:discriminators}
\emph{\ac{VT}} is an example of a tachycardia originating in the ventricles, in which the ventricles spontaneously beat at a very high rate.
If the \ac{VT} is sustained, or degenerates into \ac{VF}, it can be fatal.
A tachycardia that originates above the ventricles is referred to as a \emph{\ac{SVT}} and is a diseased but non-fatal condition.
In what follows, we will refer to sustained \ac{VT} and \ac{VF} together as \ac{VT}.
\emph{The ICD's main task is to discriminate \ac{VT} from \ac{SVT} and deliver therapy to the former only}.

\begin{figure}[t]
	\centering
	\includegraphics[scale=0.4]{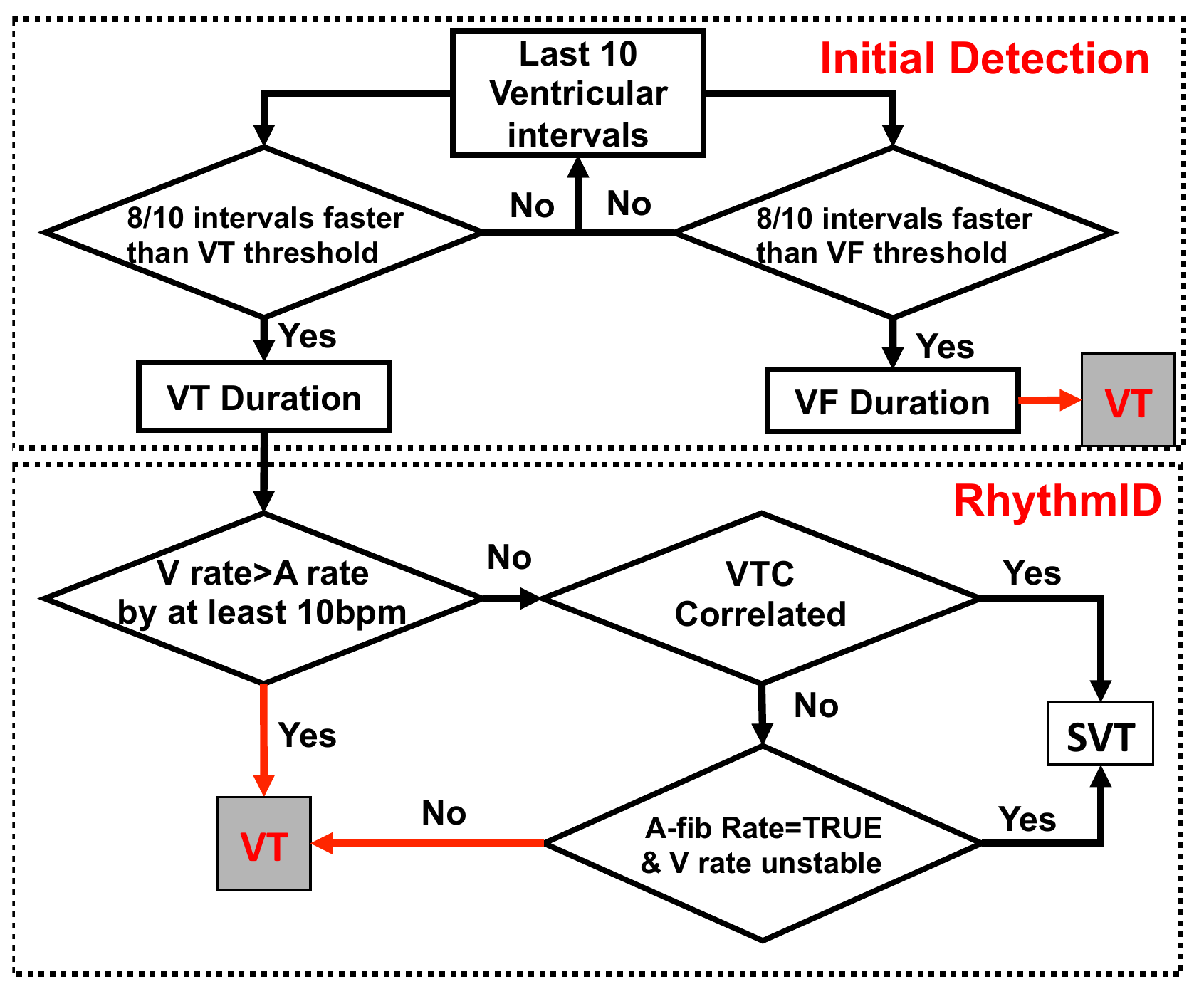}
	\vspace{-10pt}
	\caption{Boston Scientific's detection algorithm}
	\vspace{-10pt}
	\label{fig:bsc detection}
\end{figure}

Most \ac{VT}/\ac{SVT} detection algorithms found in ICDs today are composed of individual \emph{discriminators}. 
A discriminator is a software function whose task is to decide whether the current arrhythmia is SVT or VT. 
No one discriminator can fully distinguish between SVT and VT.
Thus a detection algorithm is often a decision tree built using a number of discriminators \emph{running in parallel}.
The detection algorithm of Boston Scientific is shown in Fig. \ref{fig:bsc detection} \cite{compass}.
We have modeled each discriminator in this detection algorithm as a STORMED hybrid system.
The algorithm itself is then a hybrid system.
\textbf{The ICD system is thus 
$\mathbf{\Sys_{ICD} = \Sys_{Sense}||\Sys_{Detection-Algo}}$ where $\mathbf{\Sys_{Detection-Algo}}$ is the parallel composition of the discriminator systems}.
In what follows, we present three of these discriminators we modeled, which are found in most ICDs and model them as hybrid systems, and prove they are STORMED.

\subsection{Three Consecutive Fast Intervals}
\label{sec:tcfi}
\begin{figure}[t]
	\centering
	\includegraphics[scale=0.3]{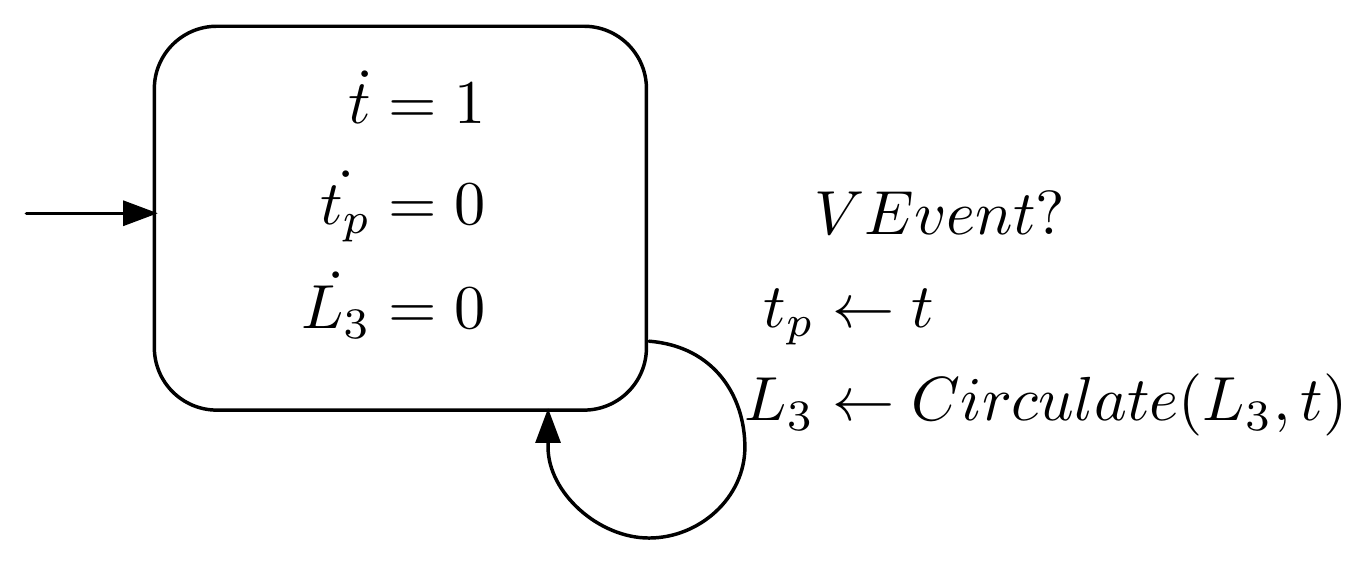}
		\vspace{-10pt}
	\caption{Three Consecutive Fast Intervals $\Sys_{TCFI}$}
	\label{fig:tcfi}
\end{figure}
Our first module simply detects whether three consecutive fast intervals have occurred, where `fast' means the interval length, measured between 2 consecutive peaks on the \ac{EGM} signal, is shorter than some pre-set amount.
See Fig. \ref{fig:tcfi}.
States $t$ and $t_p$ are clocks as before.
The vector $L_3$ is three-dimensional, and stores the values of the last three intervals.
The event VEvent? is shorthand for the transition $y(t) \geq Th$ being taken by the $\Sys_{Sense}$ automaton.
In other words, it indicates a ventricular event.
Then $L_3$ gets reset to $L_3^+ = (z_1,z_2,z_3)^+ \defeq \text{Circulate}(L_3,t-t_p)$ where
\begin{equation}
L_3^+ = 
\left(\begin{matrix}
z_2\\z_3\\t-t_{p}\\
\end{matrix}
\right)
=
\left(\begin{matrix}
0 & 1 & 0\\0& 0& 1\\0& 0 &0
\end{matrix}
\right) L_3 + 
\left(\begin{matrix}
0\\0\\t-t_p
\end{matrix}
\right)
\end{equation}
\begin{lemma}
	$\Sys_{TCFI}$ is STORMED.
\end{lemma}
\begin{prf}
We show that the reset are monotonic - the other properties are easily checked.
For reset monotonicity, we invoke the fact that there is a minimum beat-to-beat separation: heartbeats can't follow one another with vanishingly small delays. 
In other words, there exists $m>0$ such that $t - t_p^- > m$.
Similarly, there's a maximum delay between two heartbeats, call it $B$.
Now, we seek a vector $\phi \in \Re^5$ s.t. 
\begin{equation}
\label{eq:tcfi resets}
\phi \cdot \left(\begin{matrix}
t-t\\t - t_p\\L_3^+ - L_3\\
\end{matrix}
\right) = \phi_p(t-t_p) +\phi_{L_3} \cdot \underbrace{\left(\begin{matrix}
z_2-z_1\\z_3-z_2\\t-t_p - z_3\\
\end{matrix}
\right)}_{\delta} \stackrel{Want}{\geq} \zeta > 0
\end{equation} 
Now $|\delta|$ is upper bounded by $\sqrt{3\cdot (2B)^2}$ since each element is the difference of intervals shorter than $B$.
Also, $t-t_p^- > m > 0$.
So choose $\phi_{L_3} = (\phi_{z,1},\phi_{z,2},\phi_{z,3})>0$ element-wise.
\eqref{eq:tcfi resets} is satisfied if the following stronger inequality is satisfied, which can be achieved by an appropriate choice of $\phi_{z,i}$: \;
$\phi_p m \geq \zeta + \sqrt{12B^2}\sum_1^3\phi_{z,i}$
\end{prf}

\subsection{Vector Timing Correlation}
\label{sec:VTC}
\begin{figure}[t]
	\centering
	\vspace{-15pt}
	\includegraphics[scale=0.3]{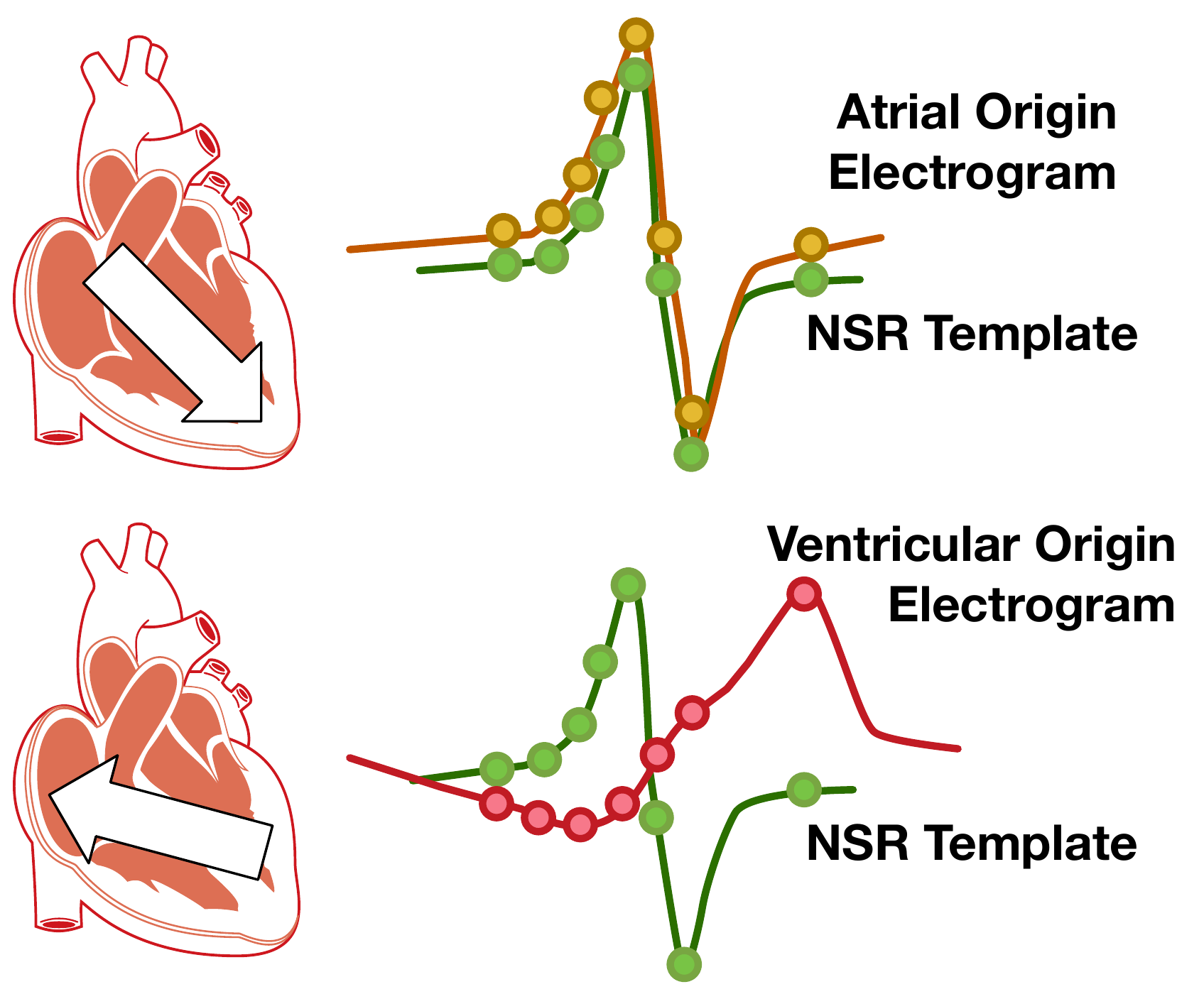}
	\caption{\small \acp{EGM} of different origin have different morphologies. The correlation of an \ac{EGM} with respect to a stored \ac{EGM} template is used to determine the origin.}
	\label{fig:egmmorphology}
	\vspace{-10pt}
\end{figure}
It has been clinically observed that a depolarization wave originating in the ventricles (as produced during \ac{VT} for example) will in general produce a different \ac{EGM} morphology than a wave originating in the atria (as produced during \ac{SVT}) \cite{compass}.
See Fig. \ref{fig:egmmorphology}.
A morphology discriminator measures the correlation between the morphology of the current \ac{EGM} and that of a stored \emph{template} \ac{EGM} acquired during normal sinus rhythm.
If the correlation is above a pre-set threshold for a minimum number of beats, then this is an indication that the current arrhythmia is supraventricular in origin.
Otherwise, it might be of ventricular origin.

Boston Scientific's implementation of a morphology discriminator is called Vector and Timing Correlation (VTC).
VTC first samples 8 \emph{fiducial} points $\egm_i,i=1,\ldots,8$ on the current \ac{EGM} $\egm$ at pre-defined time instants.
Let $\egm_{m,i}$ be the corresponding points on the template \ac{EGM}.
The correlation is then calculated as \cite{compass}
\[\rho_{new} = \frac{(8\sum_i \egm_i \egm_{m,i} - (\sum_i \egm_i)(\sum_i \egm_{m,i}))^2}{(8\sum_i \egm_i^2 - (\sum_i \egm_i)^2)(8\sum_i \egm_{m,i}^2 - (\sum_i \egm_{m,i})^2)} 
\]
Note that $\egm_m$ is a constant for the purposes of this calculation: it does not change during an execution of VTC. 
If 3 out of the last 10 calculated correlation values exceed the threshold, then \ac{SVT} is decided and therapy is withheld.

\begin{figure}[t]
\centering
\includegraphics[scale=0.325]{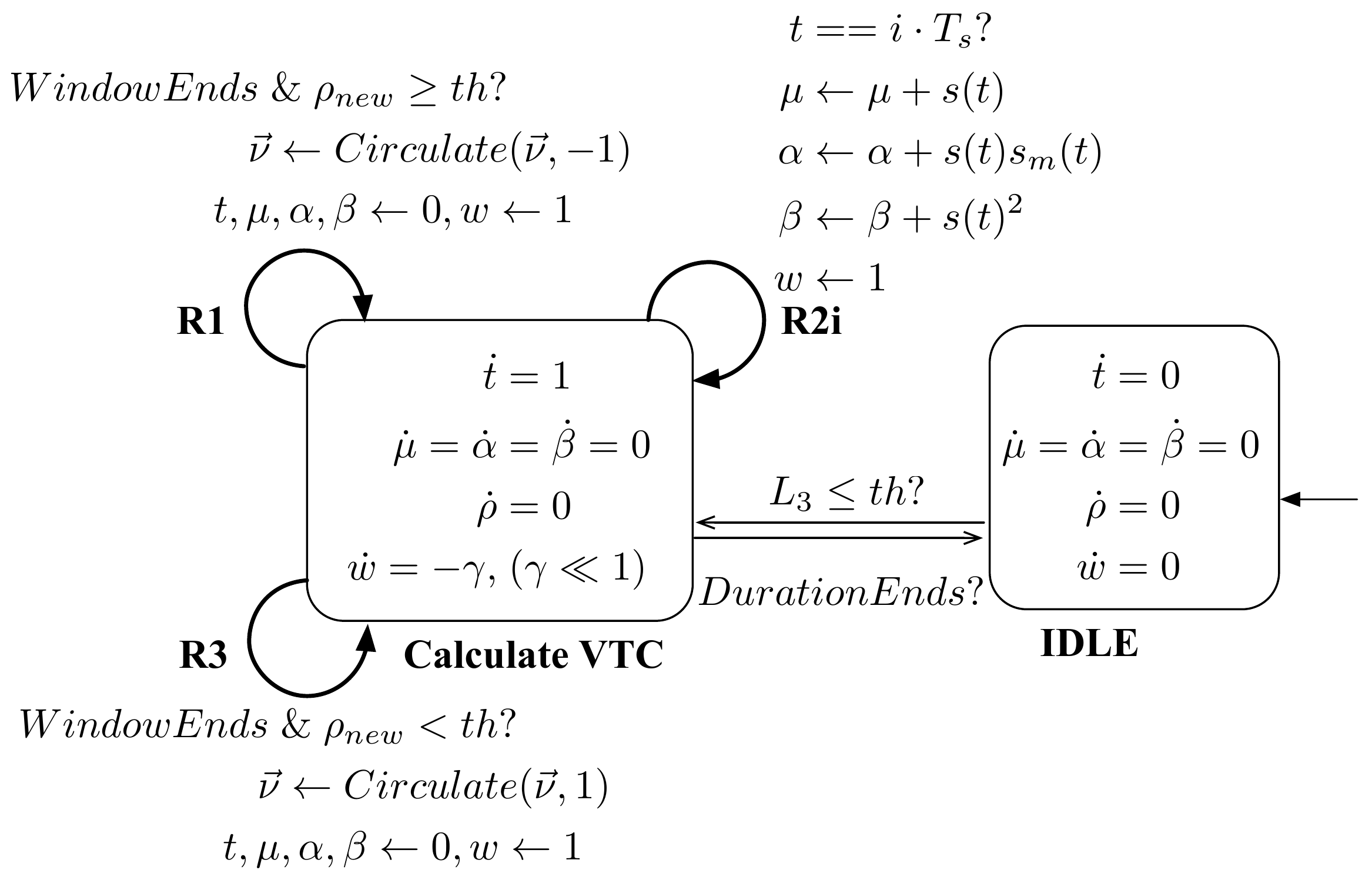}
\vspace{-10pt}
\caption{VTC calculation. $iT_s$ is the sampling time for the $i${th} fiducial point, $i=1,\ldots,8$. $R2_{1},\ldots,R2_{8}$ are the corresponding resets. For clarity of the figure, 8 transitions are represented on the same edge.}
\vspace{-10pt}
\label{fig:HVTC}
\end{figure}

The system of Fig. \ref{fig:HVTC} implements the VTC discriminator.
As before, $t$ is a local clock.
$\mu$ accumulates the values of the current \ac{EGM}, $\alpha$ accumulates the product $\egm_i \egm_{m,i}$, 
$\beta$ accumulates $\egm_i^2$.
State $w$ is an auxiliary state we need to establish the STORMED property.
$\vec{\nu}$ is a 10D binary vector: $\nu_i = -1$ if the $i^{th}$ correlation value fell below the threshold, and is $+1$ otherwise.
$L_3$ is the state of $\Sys_{TCFI}$: the guard condition $L_3 \leq th$ indicates that all its entries have values less than the tachycardia threshold, which is when $\Sys_{VTC}$ starts computing.
$WindowEnds$ indicates the `end' of an \ac{EGM}, measured as a window around the peak sensed by $\Sys_{Sense}$.  
\begin{lemma}
	\label{lemma:vtc}
	$\Sys_{VTC}$ is STORMED.
	\end{lemma}
\begin{prf}
\textbf{S}eparability obtains by observing that a uniform minimum time passes between beats and between samples. 
\textbf{T}ISG is immediate.
\textbf{O}-minimality is established by observing that all sets and functions are definable in $\Lc_{\exp}$. \textbf{ED} holds because the state space is bounded.
We now show monotonicity.
The state of the system is $x = (t , \mu ,\alpha, \beta , \vec{\nu}, w)^T \in \Re^{4+10+1}$.
Let $\phi = (\phi_c,\phi_\mu,\phi_\alpha,\phi_\beta,\phi_1,\ldots,\phi_{10},\phi_w)^T \in \Re^{15}$ be the corresponding vector.
For flows in mode CalculateVTC, we seek a $\phi$ and $\varepsilon > 0$ such that 
$\phi \cdot (t+\tau -  t , \mathbf{0} , -\gamma(t+\tau) + \gamma t) = \phi_c \tau + \phi_w(-\gamma \tau) \geq \varepsilon \sqrt{\tau^2 + \gamma^2\tau^2}$,
which is equivalent to 
$\boxed{\phi_c - \phi_w\gamma \geq \varepsilon \sqrt{1 + \gamma^2}}$.
Reset monotonicity for resets R1, R2, R3 provides three more constraints on $\phi$ and $\varepsilon$:
\begin{eqnarray*}
&\mathbf{(R1)} &
\phi \cdot (-t, -\mu , -\alpha , -\beta ,\nu_2 - \nu_1 , \nu_3 - \nu_2 , \ldots , -1 - \nu_{10}, \yhl{1-w})
\\	
&=& -\phi_c t -\phi_\mu \mu -\phi_\alpha \alpha - \phi_\beta \beta + \sum_{i=1}^{10}\phi_i(\nu_{i+1}-\nu_i) 
\\
&&+ \phi_w(1-w) \stackrel{Want}{\geq} \zeta
\\
&\mathbf{(R2)} &
\phi \cdot (t-t , \egm , \egm \egm_m, \egm^2 , \mathbf{0} , 1-w)
\\
&=& \phi_\mu \egm + \phi_\alpha \egm \egm_m+ \phi_\beta \egm^2 + \phi_w (1-w) \stackrel{Want}{\geq} \zeta
\\ 
&\mathbf{(R3)} &
 -\phi_c t -\phi_\mu \mu -\phi_\alpha \alpha - \phi_\beta \beta + \sum_{i=1}^{10}\phi_i (\nu_{i+1}-\nu_i) 
 \\
 && + \phi_w(1-w) \stackrel{Want}{\geq} \zeta
\end{eqnarray*}
where $\nu_{11} \defeq -1$ in $\mathbf{R1}$ and $\nu_{11}\defeq 1$ in $\mathbf{R3}$.
Combine $\mathbf{R1}$ and $\mathbf{R3}$ by choosing $\phi_1 = \ldots = \phi_{10}=\phi_\mu = \phi_\alpha = \phi_\beta = 0$:
\begin{eqnarray*}
\label{eq:R23}
\mathbf{(R1,3)}\; -\phi_c t + \phi_w(1-w) \geq \zeta
\\
\mathbf{(R2)}\; \phi_w(1-w) \geq \zeta
\end{eqnarray*}
Now note that when a reset occurs, $0<w \leq 1-\gamma T_s \defeq w_m$ where $T_s$ is the smallest sampling period, and that $t\leq 10B$, $B$ = the maximum peak-to-peak interval, so $\mathbf{(R2)} ,\mathbf{(R1,3)} $ can be jointly satisfied if $\boxed{-\phi_c10B + \phi_w(1-w_m) \geq \zeta}$.
The 2 boxed equations can be jointly satisfied.
	\end{prf}

\subsection{Stability discrimination}
\label{sec:stability}
\emph{Stability} refers to the variability of the peak-to-peak cycle length.
A rhythm with large variability (above a pre-defined threshold) is said to be \emph{unstable}, and is called stable otherwise.
The Stability discriminator is used to distinguish between atrial fibrillation, which is usually unstable, and \ac{VT}, which is usually stable.

\begin{figure}[t]
	\centering
	\includegraphics[scale=0.3]{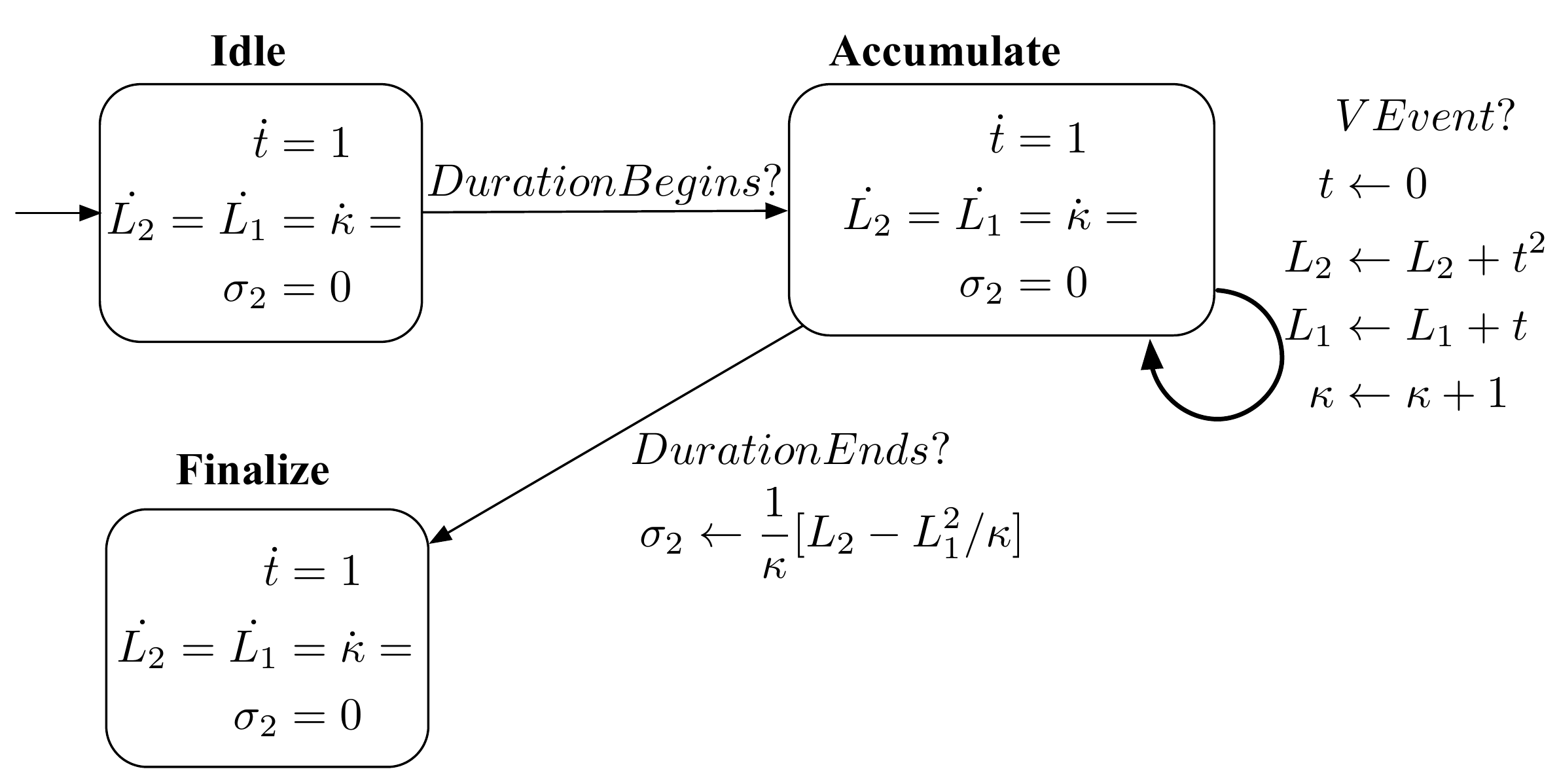}
	\vspace{-10pt}
	\caption{Stability discriminator.}
	\vspace{-10pt}
	\label{fig:Hstab}
\end{figure}

The Stability discriminator shown in Fig. \ref{fig:Hstab} simply calculates the variance of the cycle length over a fixed period called a Duration (measured in seconds).
Let $DL \geq 0$ be the Duration length.
\yhl{The events $DurationBegins?$ and $DurationEnds?$ indicate the transitions} of a simple system that measures the lapse of one Duration (not shown here).
State $t$ is a clock, $L_1$ accumulates the sum of interval lengths (and will be used to compute the average length), 
$L_2$ accumulates the squares of interval lengths,
and $\kappa$ is a counter that counts the number of accumulated beats.
$\sigma_2$ is assigned the value of the variance given by $\frac{1}{\kappa}[L_2 - L_1^2/\kappa]$

\begin{lemma}
	\label{lemma:stability}
	$\Sys_{Stab}$ is STORMED.	
\end{lemma}
The proof is in the Appendix.

Now that each system was shown to be STORMED, it remains to establish that their parallel composition is STORMED.
This result does not hold in general - Thm.~\ref{thm:SHS composition} gives conditions under which parallel composition respects the STORMED property.
Intuitively, we require that whenever a sub-collection of the systems jumps, the remaining systems that did not jump are separated from all of their respective guards by a uniform distance.
This is a requirement that can be shown to hold for our systems by modeling various minimal delays in the systems' operation. 
We may now state:
\begin{thm}
Consider the collection of systems $\Sys_{CA}$, $\Sys_{ICD} = \Sys_{Sense} || \Sys_{Detection-Algo}$ where the latter is the parallel composition of the discriminator systems.
This collection satisfies the hypotheses of Thm. \ref{thm:SHS composition} (Section \ref{sec:compositionality}) and therefore the parallel system  $\Sys_{CA} || \Sys_{ICD}$ is STORMED and has a finite bisimulation.
\end{thm}

\section{Composing STORMED systems}
\label{sec:compositionality}
The results in this section and the next apply to STORMED systems in general, including those with time-unbounded operation.
We write $[m] = \{1,\ldots,m\}$.
Given hybrid systems $\Sys_1,\ldots,\Sys_m$ in this section, $x^i, \guard^i, \theta^i,\ldots$ etc refer to a state, guard, flow $\ldots$ of system $\Sys_i$, $i\leq m$.
We show that the parallel composition of SHS is still a SHS.
Recall that $\theta_{\ell}(t;x)$ is the flow starting at $(\ell,x)$.
Given hybrid systems $\Sys_1,\ldots,\Sys_m$, their parallel composition $\Sys = \Sys_1 || \ldots ||\Sys_m$ is defined in the usual way:
$\Sys.\stSet = \Pi_i \stSet^i$,
$\Sys.\modeSet = \Pi_i \modeSet^i$,
$\Sys.\hsSet_0 =\Pi_i \hsSet_0^i$,
$Inv(\ell) = \Pi_{i}Inv^i(\ell^i)$,
$\theta_{\ell}(x,t)= [\theta_{\ell^1}^1(x^1,t)(t),\ldots,\theta_{\ell^m}^m(x^m,t)(t) ]^T$.
The system jumps if any of its subsystems jumps, so its guard sets are of the form 
$A^1\times\ldots \times A^m$ where for at least one $i$, $A^i$ is a guard of $\Sys_i$, and for the rest $A^j =\stSet^j$.
When a guard of a subsystem is satisfied, the state of that subsystem is reset according to its reset map.
The guards are made disjoint to avoid non-determinism.
\yhl{A system $\Sys$ is \emph{deterministic} if to every initial state $(\mode,\stPt)$, $\Sys$ produces a unique trajectory starting there.}

In general $\Sys$ is not separable: indeed for any candidate value of $d_{min}$, one could find a transition $(i,j)$ of $\Sys$ due to, say, a jump of $\Sys_1$, s.t. at that moment $x^2$ is closer than $d_{min}$ to one of its own guards, say $\guard^2_{(j^2,k^2)}$. 
This causes $\Sys$ to further jump $j \rightarrow k$ without having traveled the requisite minimum distance, thus violating the separability of $\reset_{ij}(\guard_{ij})$ and $\guard_{jk}$.
Therefore we need to impose an extra condition on minimum separability \emph{across} sub-systems.
\begin{thm}
	\label{thm:SHS composition}		
	Let $\SHS_i = (\Sys_i, \Ac, \phi^i,b^{i,-},b ^{i,+}, d_{min}^i, \varepsilon^i, \zeta^i)$, $i=1,\ldots,m$ be deterministic SHS 
	defined using the same underlying o-minimal structure, 
	and where each state space $\stSet^i$ is bounded by $B_{X^i}$.
	\\
	Define parallel composition $\SHS = (\Sys, \Ac, \phi,b^-,b^+, d_{min}, \varepsilon, \zeta)$ where
	$\Sys = \Sys_1 || \ldots ||\Sys_m$,	
	$\phi = (\phi^1,\ldots,\phi^m)^T \in \Re^{mn}$,
	$b^{i,-} = \inf_{x \in \stSet} \phi\cdot x$,
	$b^{i,+} = \sup_{x \in \stSet} \phi\cdot x$,
	$\varepsilon = \min(\min_i \varepsilon^i, \min_i \frac{\zeta^i}{B_{X^i}})$,
	$\zeta = \min_i \zeta^i$ and
	\[d_{min} = \min_{I\subset [m]} (\min_{i\in I}d_{min}^i, \min_{i\in I ,j \in [m]\setminus I }d_{min}^{ij})\]	
	Assume that the following \textbf{Collection Separability} condition holds: 	
	\yhl{for all $i,j \leq m, \neq j $ there exists $d_{min}^{ij}>0$ s.t. 
		\yhl{if $\stPt \in \stSet$ is in the reachable set of $\Sys$} and 
		$x^i \in G^i_e \land x^j \notin G^j_{e'} \; \forall e' \in E^j$ 
		then $d(x^j,G^j_{e'}))>d_{min}^{ij}$ for all $e'\in E^j$ 
		where $E^j$ is the edge set of $\SHS_j$ and $G^j_{e'}$ is a guard of $\SHS_j$ on edge $e' \in E^j$.}
	Then $\SHS$ is STORMED.
\end{thm}
\begin{prf}
\textbf{(S)} In $\Sys$, let $y=(y^1,\ldots,y^m)=\reset_e((x^1,\ldots,x^m))$ and assume that it was $\Sys_1$ that caused the jump.
Thus $y^j=x^j,j >1$.
Write $e=(\mode,\mode ')$.
By Collection Separability,  $d(y^j,G^j_{e^j})>d_{min}^{1j}$ for all $j>1,e^j\in E^j$, 
and by separability of $\Sys_1$ $d(y^1,G^1_{e^1}) > d_{min}^1$ for all $e^1 \in E^1$.
So by $d(y,G_{\mode', \mode ''}) > \min (d_{min}^1, \min_{j>1}d_{min}^{1j}) > d_{min}$ for any guard leading out of $\mode'$, and we have separability.
The argument can be repeated for any subset $I \subset [m]$ of systems jumping simultaneously.
\\
\textbf{(T)}:
The $\Sys$ flow $\theta_\mode(t;x)$ is TISG because the component flows $\theta^i_{\mode^i}(t;x^i)$ are TISG.
\\
\textbf{(O)} The cartesian product of definable sets is definable, so the system $\Sys$ is o-minimal.
\\
\textbf{(RM)} First we show that resets of $\Sys$ are monotonic, then that the flows of $\Sys$ are monotonic.
Let $p,q \in \modeSet$ be two modes of $\Sys$, $p\neq q$.

\underline{Case 1: $\Sys$ jumps $p \rightarrow p$}.
So any subsystem $\Sys_i$ either jumped $p^i \rightarrow p^i$ or didn't jump at all.
If $x^+ = x^-$, then (RM) is satisfied.
Else, define $\phi \defeq (\phi^1,\ldots,\phi^m) \in \Re^{n\cdot m}$, where $\phi^i$ is the $\phi$ vector of system $\Sys_i$.
Then $\phi \cdot (x^+ - x^-) = \sum_{i\in K}\phi^i\cdot(x^{i,+} - x^{i,-})$, 
where $K \subset [m]$ is the set of indices of sub-systems that jumped with $x^{i,-} \neq x^{i,+}$.
Note that $K$ depends on $x^-,x^+$.
For all $x^-,x^+$ pairs (and so for all $K$) 
$\sum_{i\in K} \zeta^i \geq \min_{i \in [m]}\zeta^i \defeq \zeta > 0$.
So by (RM) for each $\Sys_i$,
\begin{equation*}
\label{eq:parallel rm1}
\phi \cdot (x^+ - x^-) =  \sum_{i\in K}\phi^i\cdot(x^{i,+} - x^{i,-}) \geq \sum_{i\in K}\zeta^i \geq \zeta > 0
\end{equation*}
Thus (RM) is satisfied.

\underline{Case 2: $\Sys$ jumps $p \rightarrow q$}.
At least one syb-system $\Sys_i$ jumped $p^i \rightarrow q^i \neq p^i$.
Then 
$\phi \cdot (x^+ - x^-) = \sum_{i\in [m]}\phi^i\cdot(x^{i,+} - x^{i,-}) = \sum_{i\in K}\phi^i\cdot(x^{i,+} - x^{i,-})$,
where $K = K_= \cup K_{\neq} \subset [m]$ and 
$K_=$ is the index set of subsystems that jumped $p^i \rightarrow p^i$ with $x^{i,+} \neq x^{i,-}$, 
and $K_{\neq}$ is the index set of subsystems that jumped $p^i \rightarrow q^i \neq p^i$ with $x^{i,+} \neq x^{i,-}$.
Subsystems that didn't jump or jumped without changing their continuous state don't contribute to the sum.
Note that $K_=,K_{\neq}$ depend on $x^-,x^+$.
So we have
$\phi \cdot (x^+ - x^-) \geq \sum_{i\in K_{\neq}}\varepsilon^i ||x^{i,+} - x^{i,-}|| + \sum_{i\in K_=} \zeta^i$.

For all $X^i$, $||x^{i,+} - x^{i,-}|| \leq B_{X^i}$ for all $x^{i,-},x^{i,+} \in X^i$.
Therefore $\zeta^i \frac{||x^{i,+} - x^{i,-}||}{B_{X^i}} \leq \zeta^i$ for all $i\in K$.
So 
\begin{eqnarray*}
&&	\phi \cdot (x^+ - x^-) \geq
\\ 
&\;&\sum_{i\in K_{\neq}} (\min_{i\in [m]}\varepsilon^i) ||x^{i,+} - x^{i,+}|| + \sum_{i\in K_=} \frac{\zeta^i}{B_{X^i}}||x^{i,+} - x^{i,-}|| \geq 
\\
&& \sum_{i\in K_{\neq}} (\min_{i\in [m]}\varepsilon^i) ||x^{i,+} - x^{i,-}|| + \sum_{i\in K_=} (\min_{i\in [m]}\frac{\zeta^i}{B_{X^i}})||x^{i,+} - x^{i,-}||
\end{eqnarray*}
Let $\varepsilon \defeq \min(\min_i \varepsilon^i, \min_i \frac{\zeta^i}{B_{X^i}})$.
Then 
\begin{equation*}
\phi \cdot (x^+ - x^-) \geq \sum_{i\in K} \varepsilon ||x^{i,+} - x^{i,-}|| \geq \varepsilon ||x^+-x^-||
\end{equation*}
So $\Sys$ has monotonic resets.

%
The flows of $\Sys$ are also monotonic along $\phi$.
Indeed for any $q \in \modeSet$,
$\phi\cdot (\theta_q(t+\tau;x) - \theta_q(t;x) ) = \sum_{i=1}^{m}\phi^i \cdot (\theta^i_{q^i}(t+\tau;x^i) - \theta^i_{q^i}(t;x^i)) \geq \sum_i \varepsilon^i ||(\theta^i_{q^i}(t+\tau;x^i) - \theta^i_{q^i}(t;x^i))|| \geq \varepsilon ||(\theta_{q}(t+\tau;x) - \theta_{q}(t;x))||$

\textbf{(ED)} By Prop. \ref{prop:ED}.
	\end{prf}
\section{Finite simulation for STORMED systems}
\label{sec:simulationAprox}
In general it is not possible to compute the reach sets required in Alg.~\ref{algo:bisimulation} exactly unless the underlying o-minimal theory is decidable.
The $\Sys_{ICD}||\Sys_{CA}$ closed loop is definable in $\Lc_{\exp}$, and the latter is not known to be decidable.\\
The authors in \cite{PrabhakarVVD09_toklerant} proposed approximating the flows and resets by polynomial flows and resets in the decidable theory $\Lc_\Re$.
However, the approximation process is typically iterative and requires manual intervention, or is restricted to subclasses of STORMED systems \cite{PrabhakarVVD09_toklerant}.

Here we show that if an approximate reachability tool with definable over-approximations is available for the continuous dynamics, it can be used in Algo~\ref{algo:bisimulation} (instead of exact reachability) to yield a finite simulation (rather than a bisimulation).
Intuitively, the additional intersections of approximate reach sets with blocks of $Q/\sim$ do not destroy finiteness of the procedure.
Since we only have a simulation, counter-examples on the abstraction should be validated in a CEGAR-like fashion.
%
\begin{lemma}
	\label{lemma:finite simu}
	Let $\SHS = (\Sys,\ldots)$ be a SHS and $\sim$ and equivalence relation on $\stSet$.
	For any mode $\mode$ of $\Sys$, its dynamical sub-system $\Dc$ with state space $X = \Sys.\stSet$ and flow $\theta_\mode $ admits a finite simulation $\simu_\mode$ that respects $\sim$, returned by Alg.~\ref{algo:bisimulation}.
\end{lemma}
The proof is in the Appendix.
Let $\Ft^\epsilon(\partition) \defeq \cap_{\mode}\simu_{\mode \in \modeSet}$ where $\partition = \stSet/\sim$. $\Ft^\varepsilon$ refines all the $\simu_\mode$'s, and it is a finite simulation of $\Sys$ by itself w.r.t. the continuous transition $\trans{\tau}$.
It is clear that $\Ft^\epsilon(\cdot)$ is idempotent: $\Ft^\epsilon(\Ft^\epsilon(\partition)) = \Ft^\epsilon(\partition)$

\begin{thm}
	\label{thm:finite simulation}
	Let $\Sys$ be a STORMED hybrid system, 
	and $\partition$ be a finite definable partition of its state space.
	Define 
	\begin{equation}
	\label{eq:Fte,Fde}
W_0 = \Ft^\epsilon(\partition), \quad \forall i\geq 0, W_{i+1} = \Ft^\epsilon(\Fd(W_i))
	\end{equation}
Then there exists $U \in \Ne$ s.t. $W_{U+1} = W_U$ and $\Ft^\epsilon(W_U)$ is a simulation of $\Sys$ by itself.
\end{thm}

\begin{prf}
By Lemma 10 of \cite{VladimerouPVD08_STORMED} there exists a uniform bound $U$ on the number of discrete transitions of any execution of the STORMED system $\Sys$, so $\Fd(W_k) = W_k$ for all $k\geq U$.
Moreover 
$W_{U+1} = \Ft^\epsilon(\Fc_d(W_U)) = \Ft^\epsilon(W_U)$
and $W_{U+2}= \Ft^\epsilon(\Fc_d(W_{U+1}))= \Ft^\epsilon(W_{U+1})= \Ft^\epsilon(\Ft^\epsilon(W_U)) = \Ft^\epsilon(W_U) = W_{U+1}$, so the iterations reach a fixed point.
The fact that $\Ft^\epsilon(W_U)$ is a simulation then yields the desired result.
\end{prf}
%
\subsection{Example: SpaceEx reachable sets}
\label{sec:spaceex}
Lemma \ref{thm:finite simulation} required that the over-approximation sets $\Rc^\epsilon_{t}(\{{x}\})$ be definable for every $x$ and $t$ (see proof).
In practice, we need to show that the over-approximation \emph{actually computed by the reachability tool} (which may not be the full ball $\Rc^\epsilon_{t}(x)$) is definable.
In this section we show that the over-approximations computed by SpaceEx \cite{FrehseCAV11} are definable.
Given the set $X\subset \Re^n$ and finite $\Vc \subset \Re^n$, parameter $\lambda \in [0,1]$ a time step $\delta>0$, and $(i,j) \in E$, 
SpaceEx over-approximates $\reset_{ij}(X)$ by $\Kc(\Vc,X) \defeq \reset_{ij}(TH_\Vc(X)\cap G_{ij})\cap Inv(j)$ and $\Rc_{\lambda \delta}^\epsilon(X)$ by \cite{FrehseCAV11}:
\begin{eqnarray}
\Omega_\lambda(X,\delta) &=& (1-\lambda)X \oplus e^{\delta A} X 
\nonumber \\
&\oplus&(\lambda E_\Omega^+(X,\delta) \cap (1-\lambda) E_\Omega^-(X,\delta))
\end{eqnarray}
where
$TH_\Vc(X) \defeq \{x\in\Re^n \;|\;\land_{\vec{a} \in \Vc} \vec{a}\cdot x \leq \rho(\vec{a},X)\}$ is the template hull of $X$ and $\rho$ its support function,
$E_\Omega^+ = \boxdot (\Phi_2 \boxdot(A^2 X)$,
$E_\Omega^- = \boxdot (\Phi_2 \boxdot(A^2 e^{\delta A}X))$,
$\oplus$ is the Minkowski sum,
$\boxdot S =  [-\overline{|x_1|}, \overline{|x_1|} ] \times \ldots \times [-\overline{|x_n|}, \overline{|x_n|} ]$ is the box hull
with $\overline{|x_i|} \defeq \max\{|x_i| \text{ s.t. } x=(x_1,\ldots,x_n) \in S\}$.

\begin{thm}
	\label{thm:spaceex definable}
	For all definable polytopes $X \subset \Re^n$, the sets $\Kc(\Vc,X)$ and $\Omega_\lambda(X,\delta)$ is definable are $\Lc_{\exp}$.
\end{thm}
	
\begin{prf}
Let $S, Y \subset \Re^n$ be two definable sets in some o-minimal structure $\Ac$.
Let $\lambda \in \Re$ and let $A$ be a real matrix.
Then the following sets are also o-minimal: $\lambda S$, $A S$, $S \cap Y$, $S \oplus Y$, $S \cap Y$, $TH_\Vc(S)$ and $\boxdot S$.
Now the result follows by noting that $\Kc(\Vc,X)$ and $\Omega_\lambda(X,\delta)$ are constructed by composing the above definability-preserving operations.
\end{prf}

\section{Conclusion}
In this paper, we presented the first formalization of a hybrid system model of the human heart and \ac{ICD} closed loop and showed that it admits a finite bisimulation, and that definable approximate reachability yields a finite simulation for STORMED systems.
\bibliographystyle{abbrv}
\bibliography{HSCC2015_CompositionalConf,houssam,fainekos_bibrefs,hscc2016,biblio2}  
%
\appendix
%
{\large \textbf{Proof of Lemma \ref{lemma:stability}.}}
\begin{prf}
We show the resets are monotonic - the other properties are immediate.
The state is $x = (t,L_2,L_1,\kappa,\sigma_2)^T$.
The self-transition ACCUMULATE $\rightarrow$ ACCUMULATE is initiated by VEvent (ventricular peak).
At reset time, $0 \leq t \leq DL$, we have that 
$\phi\cdot(0-t,t^2,t,1,0)^T \geq -\phi_1 DL + \phi_4 \stackrel{Want}{\geq} \zeta$.

The transition ACCUMULATE $\rightarrow$ FINALIZE, initiated at the end of Duration, saves the value of the variance in $\sigma_2$.
This reset produces the constraint
$\phi_5 ((L_2 -L_1^2/\kappa)/\kappa) \geq \varepsilon |((L_2 -L_1^2/\kappa)/\kappa)|$.
But the quantity in absolute value is itself a variance and so is positive, therefore the constraint is simply $\phi_5 \geq \varepsilon$, compatible with the previous inequality.
\end{prf}
{\large \textbf{Proof of Lemma \ref{lemma:finite simu}}.}
\begin{prf}
	This follows the lines of the elegant proof of \cite{BrihayeM05_ominimal} as formulated in \cite{tabuada} and generalizes it to set-valued maps.
	(The fact that using an approximate $Post$ operator yields a simulation is a special case of a more general result on transition systems but we prove it here for completeness. 
	\yhl{Also note that this result holds for o-minimal systems \cite{LaFerrierePS00_Ominimal} generally, not just STORMED systems}).
	
	First observe that using approximate reachability on a system $\Sys$ is tantamount to replacing $\Sys$ with a system $\Sys^\varepsilon$ whose flows and reset maps are set-valued $\varepsilon$ over-approximations of the flows and resets of $\Sys$ (but is otherwise unchanged).
	Therefore define the dynamical system $\Dc^\varepsilon$ with state space $\stSet$ and whose flow $\Theta: \Re \times \Re^n \rightarrow 2^{\Re^n}$ is a set-valued $\varepsilon$ over-approximation of $\theta_\mode$:
	$\Theta(t;x) = \{y \in \Re^n \;|\; ||y-\theta(t;x)||^2 \leq \epsilon^2\}$.	
	Let $\partition \defeq \stSet/\sim$ be the partition induced by $\sim$.
	It follows from the definability of $\theta$ and $||\cdot||^2$ that $\Theta$ is definable. 
	Given $P \in \partition$, let $Z(P) = \Theta^{-1}(P) \defeq \{(x,t) \;|\; \Theta(x,t) \cap P \neq \emptyset\}$.
	Then $Z(P)$ is definable because $P$ and $\Theta$ are definable.
	Let $Z_x(P) = \{t \;|\; (x,t) \in Z(P)\} \subset \Re$ be the \emph{fiber} of $Z$ over $x$.
	The number of connected components of $Z_x(P)$ equals the number of times that $\Theta(x,t)$ intersects $P$.
	Now it follows from \cite{tabuada} Thm.7.11 that there exists a uniform upper bound on the number of connected components of $Z_x(P)$, independent of $x$.
	Let that bound be $V_P$.
	Thus $\Theta(x,t)$ visits $P$ at the most $V_P$ times, regardless of $x$.
	Since there is a finite number of blocks $P \in \partition$, then $\Theta(x,t)$ visits any block $P$ a maximum of $V \defeq \max_P(V_P)$ times.
	
	Thus we can associate to each $x\in \stSet$ a finite number of finite strings $q(x) = (\ell_1,\ell_2,\ldots,\ell_{i-1},\widehat{\ell_i},\ell_{i+1},\ldots,\ell_s)$, where $\ell_i,\widehat{\ell}_i \in \partition$.
	Each $q(x)$ gives the sequence of blocks that $\Theta(x,t)$ visits (with repetition), and in which $\widehat{\ell_i}$ is the block containing $x$.
	There may be more than one such string because the set $\Theta(x,t)$ might intersect more than one block of $\partition$ at a time.		
	The length of $q(x)$ is thus uniformly upper-bounded by $V\cdot |\partition|$, so there's a finite number of different strings $q(x)$. 
	Let $\Qc(x)$ be the set of such strings associated to $x$, and let $\Qc = \cup_x \Qc(x)$.
	Then $\Qc$ is the state space of the finite transition system $K = (\Qc,\{*\},\trans{},\Qc_0)$ whose transition relation is 
	\begin{compactitem}
		\item $\ell_1\ldots\widehat{\ell}_i\ldots\ell_s \trans{*} \ell_1\ldots\widehat{\ell}_{i+1}\ldots\ell_s$
		\item $\ell_1\ldots\ell_{s-1} \widehat{\ell_s} \trans{*} \ell_1\ldots\ell_{s-1} \widehat{\ell}_s$
	\end{compactitem}

	It is clear that $K$ is non-deterministic and simulates $\Dc$ but is not a bisimulation because of the over-approximation produced by $\Theta$.	 
\end{prf} 

\end{document}